\documentclass[10pt,journal,compsoc]{IEEEtran}

\usepackage{array}
\usepackage{textcomp}
\usepackage{stfloats}
\usepackage{verbatim}
\usepackage{graphicx}
\usepackage{cite}

\usepackage{amssymb}
\usepackage{epstopdf}
\usepackage{gensymb}
\usepackage{color}
\usepackage{amsmath}
\usepackage{bm}
\usepackage{etoolbox}
\usepackage{booktabs}
\usepackage{multirow}
\usepackage{comment}
\usepackage{subcaption}
\usepackage[ruled,vlined]{algorithm2e}
\usepackage{algorithmicx}
\usepackage{mathrsfs}
\usepackage{algpseudocode}
\usepackage{hyperref}
\usepackage{float}
\usepackage{balance}

\usepackage{threeparttable}
\usepackage{mathtools}
\usepackage{hyperref}
\usepackage{pifont}
\usepackage{listings}
\usepackage[utf8]{inputenc}
\usepackage[english]{babel}
\usepackage{amsthm}
\usepackage{soul}

\hypersetup{
    colorlinks=true,
    linkcolor=black,
    filecolor=black, 
    citecolor=black,
    urlcolor=cyan,
}
\definecolor{myblue}{rgb}{0,0,0.0}
\usepackage[framemethod=TikZ]{mdframed}
\alglanguage{pseudocode}
\algnewcommand{\Initialize}[1]{%
  \State \textbf{Initialize:}
  \Statex \hspace*{\algorithmicindent}\parbox[t]{.8\linewidth}{\raggedright #1}
}

\usepackage{url}
\urldef{\mailsa}\path|{alfred.hofmann, ursula.barth, ingrid.haas, frank.holzwarth,|
	\urldef{\mailsb}\path|anna.kramer, leonie.kunz, christine.reiss, nicole.sator,|
	\urldef{\mailsc}\path|erika.siebert-cole, peter.strasser, lncs}@springer.com|    

\theoremstyle{definition}

\definecolor{R}{RGB}{0,0,150}

\theoremstyle{remark}

\ifCLASSINFOpdf
\usepackage{xspace}
\newcommand{\name}{\texttt{TRAIT}\xspace}

\usepackage{minibox}
\newcounter{observcntr}
\newcommand*{\observ}[1]{%
    \stepcounter{observcntr}%
    \begin{center}
    \vspace{-4pt}
    \minibox[frame, rule=1pt,pad=3pt]{
        \begin{minipage}[t]{0.95\columnwidth}
        \textbf{Observation~\arabic{observcntr}:} \textit{#1}.
        \end{minipage}
    }
    \vspace{-4pt}
    \end{center}
}

\usepackage{minibox}
\newcounter{takeaway}

\newcommand*\circled[1]{\tikz[baseline=(char.base)]{
            \node[shape=circle,draw,inner sep=0.7pt] (char) {#1};}}
\begin{document}

\title{From Pixels to Trajectory: \\ Universal Adversarial Example Detection via Temporal Imprints}

\author{

Yansong Gao, Huaibing Peng, Hua Ma, Zhiyang Dai, \\ Shuo Wang\IEEEauthorrefmark{2}, Hongsheng Hu, Anmin Fu, Minhui Xue


\IEEEcompsocitemizethanks{\IEEEcompsocthanksitem \IEEEauthorrefmark{2} Corresponding author.}

\IEEEcompsocitemizethanks{\IEEEcompsocthanksitem Y.~Gao is with the School of Computer Science and Software Engineering, The University of Western Australia, Australia.
garrison.gao@uwa.edu.au}

\IEEEcompsocitemizethanks{\IEEEcompsocthanksitem H.~Peng, Z.~Dai and A.~Fu are with the School of Computer Science and Engineering, Nanjing University of Science and Technology, China. \{paloze;dzy;fuam\}@njust.edu.cn}

\IEEEcompsocitemizethanks{\IEEEcompsocthanksitem H.~Ma and M.~Xue are with Data61, CSIRO, Australia. \{mary.ma;jason.xue\}@data61.csiro.au}

\IEEEcompsocitemizethanks{\IEEEcompsocthanksitem S.~Wang is with the School of Cyber Science and Engineering, Shanghai Jiao Tong University, China. wangshuosj@sjtu.edu.cn}

\IEEEcompsocitemizethanks{\IEEEcompsocthanksitem H.~Hu is with the School of Information and Physical Sciences, The University of Newcastle, Australia. hongsheng.hu@newcastle.edu.au}
}





\IEEEtitleabstractindextext{
\begin{abstract}
For the first time, we unveil discernible temporal (or historical) trajectory imprints resulting from adversarial example (AE) attacks. Standing in contrast to existing studies all focusing on \textit{spatial (or static) imprints within the targeted underlying victim models}, we present a fresh temporal paradigm for understanding these attacks.
Of paramount discovery is that these imprints are encapsulated within a single loss metric, spanning \textit{universally} across diverse tasks such as classification and regression, and modalities including image, text, and audio. 
Recognizing the distinct nature of loss between adversarial and clean examples, we exploit this temporal imprint for AE detection by proposing \name (\underline{Tr}aceable \underline{A}dversarial Temporal Trajectory \underline{I}mprin\underline{t}s). \name operates under minimal assumptions without prior knowledge of attacks, thereby framing the detection challenge as a one-class classification problem. However, detecting AEs is still challenged by significant overlaps between the constructed synthetic losses of adversarial and clean examples due to the \textit{absence of ground truth for incoming inputs}. \name addresses this challenge by converting the synthetic loss into a spectrum signature, using the technique of Fast Fourier Transform to highlight the discrepancies, drawing inspiration from the temporal nature of the imprints, analogous to time-series signals.
Across 12 AE attacks including SMACK (USENIX Sec'2023), \name demonstrates consistent outstanding performance across comprehensively evaluated modalities (image, text, audio), tasks (classification and regression), datasets (nine datasets), and model architectures (e.g., ResNeXt50, BERT, RoBERTa, AudioNet). In all scenarios, \name achieves an AE detection accuracy exceeding 97\%, often around 99\%, while maintaining a false rejection rate of 1\%. \name remains effective under the formulated \textit{strong adaptive attacks}.
\end{abstract}

}

\maketitle

%

\section{Introduction}\label{sec:Intro}
Despite the exceptional performance of deep learning (DL) models, they confront threats from adversarial attacks~\cite{mink2023security,papernot2018sok}. These attacks can result in unexpected behaviors such as adversarial example generation or backdoor implantation, or lead to privacy breaches in the underlying model or data such as through privacy inference attacks~\cite{salem2023sok,gao2020backdoor,cao2023fedrecover,liu2022membership}. 
The adversarial example (AE) attack has been a persistent security risk noted nearly a decade ago~\cite{szegedy2013intriguing}, coinciding with the rapid expansion of applications relying on deep learning. 
An adversary crafts an adversarial example by subtle perturbations to a benign example while maintaining semantic similarity between them, for example, making the alterations imperceptible to humans when observing an adversarial image example. Initially demonstrated against image data types and classification tasks, AEs have extended their impact to various other data types such as audio~\cite{abdullah2021sok,hussain2021waveguard}, text~\cite{gao2018black,li2018textbugger}, graphs~\cite{xu2020adversarial,dai2018adversarial}, and non-classification tasks~\cite{balda2019perturbation,xie2017adversarial,song2018physical,zhang2023capatch}. Consequently, this poses a significant threat to the security and safety of DL applications. These threats include breaching facial recognition~\cite{an2023imu}, speech recognition~\cite{abdullah2021sok} to gain unauthorized access, manipulation of self-driving systems~\cite{255240}, tampering with pedestrian detection~\cite{wang2023does}, generating harmful text~\cite{menczer2023addressing}, evading malware and spam detection~\cite{chen2023obsan}, and bypassing image safety checks in social networks to perpetuate cyberbullying~\cite{bethany2023towards}.

\noindent{\bf Limitations of State-of-the-Art.} 
Though significant strides have been made in developing countermeasures against such attacks, as outlined in the related work (\autoref{sec:relatedwork}), current methods still face three notable limitations when it comes to neutralizing specific threats.

\noindent$\bullet$ \textit{Limitation 1: Limited Prior Knowledge or Unseen Attacks}. 
A major limitation in current adversarial example defense approaches is their dependency on prior knowledge of AE techniques~\cite{ma2019nic,zhu2023ai,shan2020gotta}, especially evident in adversarial training and detection strategies. These methods often utilize insights from previously encountered AEs to enhance model robustness. While it improves resilience against known or similar attack patterns, its efficacy diminishes significantly in the face of novel, unseen adversarial threats (especially adaptive attacks), highlighting a crucial limitation in defense without prior AE knowledge.

\noindent$\bullet$ \textit{Limitation 2: Maintaining Primary Task Performance}. 
Enhancing DL model resilience through adversarial training often necessitates alterations in the training procedure or model architectures, diverging from conventional settings~\cite{rade2021reducing,madry2018towards,sehwag2021robust,zhu2023towards}. Such modifications can introduce significant computational demands and potentially compromise the model's performance on legitimate examples. Moreover, certain defense strategies that involve input preprocessing, like those requiring GANs~\cite{yang2022you,xu2017feature,bethany2023towards} or diffusion models~\cite{zhang2023diffsmooth}, face challenges related to training stability and high resource consumption, raising concerns about the practicality and efficiency of these approaches in preserving the primary task's fidelity. This limitation in existing works demands a solution without undermining the primary task performance.

\noindent$\bullet$ \textit{Limitation 3: Generalization cross Modalities or Tasks}. 
A significant portion of existing AE defense strategies, particularly those based on input transformations and detection mechanisms, are developed with a focus on image data, limiting their applicability across other modalities such as text and audio~\cite{yang2022you,xu2017feature,ho2022disco,xiang2022patchcleanser, wang2023addition}. Additionally, there is a notable lack of defense methods capable of extending their protective measures beyond classification tasks to include other forms such as regression~\cite{yang2022you,ma2019nic,zhu2023towards}, highlighting the need for a more universally applicable solution.

\begin{figure}[t]
    \centering
    \includegraphics[trim=0 0 0 0,clip,width=0.35 \textwidth]{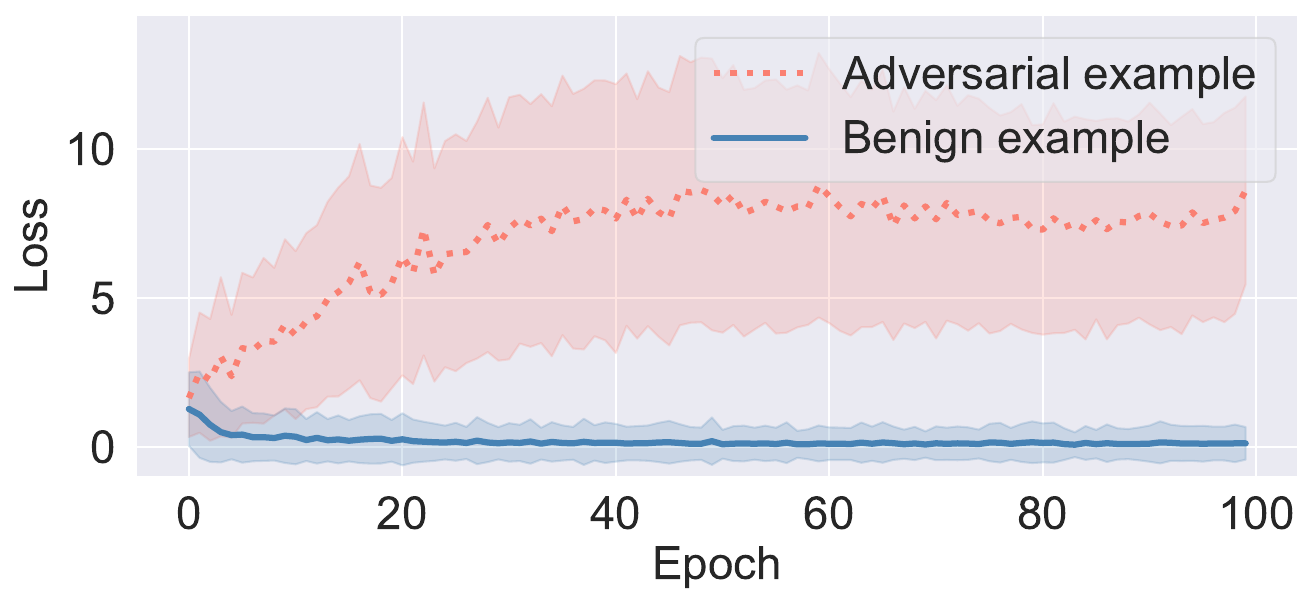}
        \caption{Losses of 1,000 clean and (untargeted) AEs on 100 intermediate models (CIFAR10+ResNet18). AE attack method is FGSM with $\epsilon = 8/255$ under $\ell_\infty$-norm. Note the loss here is computed with the \textit{ground-truth} label.}
    \label{fig:FGSM_loss}
\end{figure}

Upon these prohibitive limitations, we throw the following research question:
\begin{mdframed}[backgroundcolor=black!10,rightline=false,leftline=false,topline=false,bottomline=false,roundcorner=2mm]
Can we develop a universally applicable method for detecting AEs that operates efficiently and effectively across various modalities and tasks, independent of prior knowledge about specific attacks?
\end{mdframed}

\noindent{\bf Technical Challenges.} Upon examining existing literature (elaborated in \autoref{sec:relatedwork}), we recognize that a solution to resolve the above research question is immensely challenging, perhaps even seemingly insurmountable. Traditional defense mechanisms primarily focus on spatial or static features within the model itself ({i.e., they harness information from the \textit{deployed underlying model solely} to detect AEs}), neglecting the rich temporal information intrinsically available through the model's training history. 
To address this challenging research question, it is important to resort to a \textit{new paradigm} that leverages the temporal trajectories imprinted by AEs during the model's training phases. 
Model training involves multiple epochs, each producing an intermediate model (IM). While the final model is often used in deployment, these IMs, reflective of the model's learning journey, are usually discarded. For the first time, we highlight that AEs leave distinct, traceable imprints within these temporal trajectories. When an AE is introduced to these IMs, its trajectory deviates from benign input. 
However, tapping these temporal imprints for AE detection still faces several technical challenges (\textbf{Cs}).

\noindent\textbf{C1: Universal Metric for Trajectory Imprints.} Identifying a metric that consistently reflects trajectory imprints universally across different tasks and modalities is challenging. Conventionally used metrics like latent representations, logits, and softmax values, while informative, are high-dimensional and may become unwieldy, especially in scenarios with a vast number of categories like facial recognition. Furthermore, their applicability can be limited by model architecture and task specificity, rendering some metrics, such as softmax values, unsuitable in contexts like regression tasks.

\begin{figure}[t]
    \centering
    \includegraphics[trim=0 0 0 0,clip,width=0.40 \textwidth]{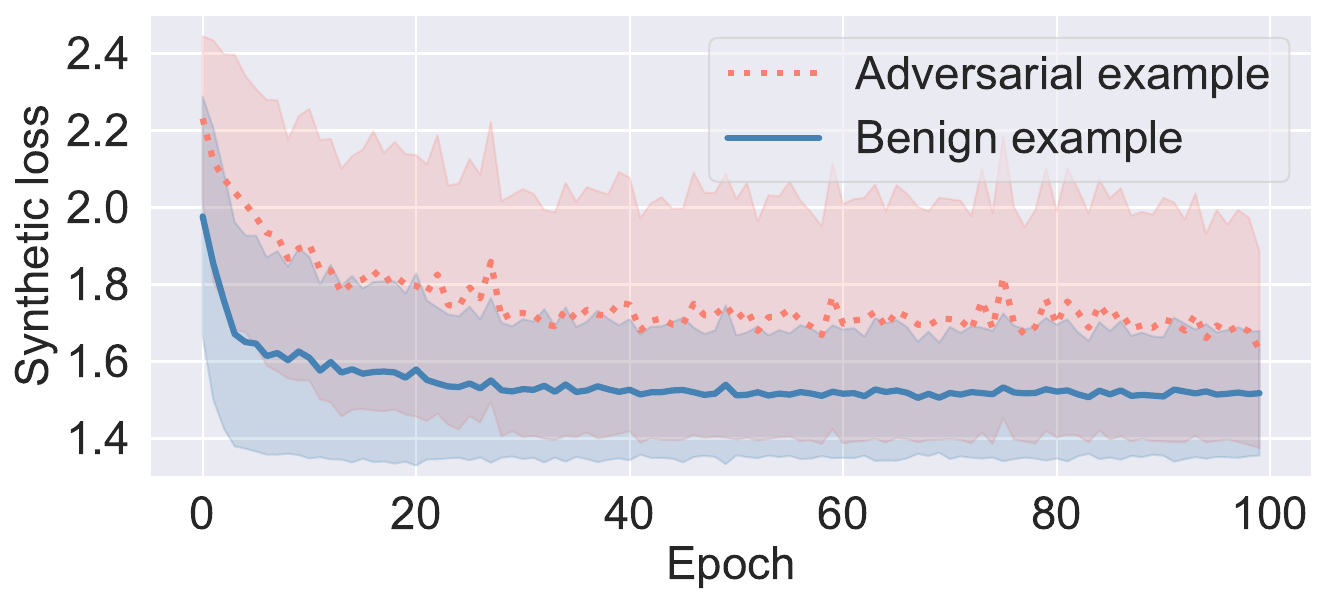}
        \caption{Synthetic losses of 1,000 clean and (untargeted) adversarial examples, respectively, on 100 intermediate models. Other settings are the same to~\autoref{fig:FGSM_loss}.}
    \label{fig:FGSM_fake_loss}
\end{figure}

\begin{figure}[t]
    \centering
    \includegraphics[trim=0 0 0 0,clip,width=0.50 \textwidth]{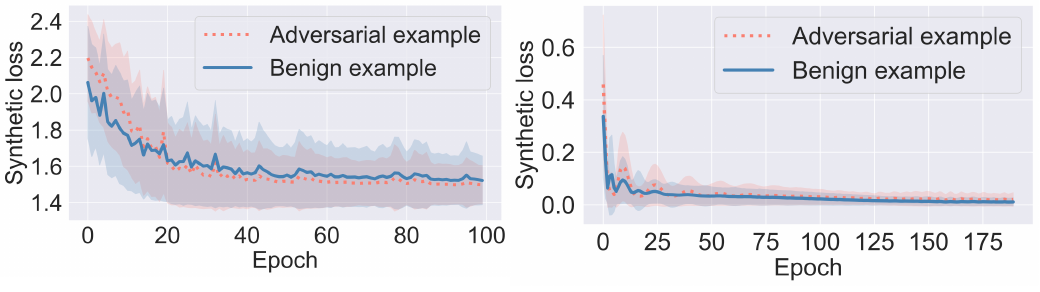}
        \caption{Synthetic loss of (left) classification task of STL10 with ResNet18, and (right) multivariate time-series regression task, temperature forecasting with LSTM.}
    \label{fig:FGSM_overlap}
\end{figure}

\noindent{\bf C2: Distinguishability of Overlapping Loss Imprints.} The task of differentiating benign inputs from AEs becomes particularly challenging due to the substantial overlap observed in their loss trajectories over the model's training epochs. This overlap, as illustrated in \autoref{fig:FGSM_fake_loss} and \autoref{fig:FGSM_overlap}, makes it difficult to identify adversarial interventions. The root of this challenge is the \textit{absence} of explicit ground-truth for online incoming data, which would otherwise provide a clear benchmark for comparison. Without this definitive reference, the loss imprints of adversarial and benign examples appear deceptively similar, complicating the detection process. 

\noindent\textbf{C3: Navigating Uncharted Attacks.}
A recognized challenge in adversarial defense is the absence of prior knowledge about new or evolving adversarial techniques. This lack of insight into the continually advancing landscape of adversarial strategies poses a significant barrier to creating defenses that are resilient to evolving threats. Traditional defense mechanisms often rely on known attack patterns to train and fortify models, leaving them vulnerable to unseen attacks, especially adaptive attacks.

\noindent{\bf Our Solution.} Through a meticulous examination of the outlined challenges, we have developed a framework that proficiently identifies AEs, seamlessly extending its capabilities across diverse data modalities, tasks, and model architectures with preserved efficiency and effectiveness. 

To address the first challenge \textbf{C1}, we resort to the loss as a universal metric to quantify the trajectory imprint. The loss metric stands independent of model architecture, task and modality, offering a standardized universal measure to characterize the trajectory imprint, circumventing the complexities associated with high-dimensionality and task-specific dependencies. As exemplified in~\autoref{fig:FGSM_loss}, it displays the losses of 1,000 adversarial and clean examples across about 100 IMs where the \textit{ground-truth is assumed to be known}. We can see that the loss curves are different between them. 
The significant disparity in the loss imprint between adversarial and benign examples appears promising. However, a critical factor to note is the absence of ground-truth for incoming inputs, making direct loss computation impossible. We overcome it via \textit{synthetic loss}, where the predictions from the target model serve as the surrogate ground-truth (as elaborated in \autoref{sec:synthetic_loss}). However, synthetic loss poses the challenge \textbf{C2}, where the loss imprints of adversarial and clean examples become substantially overlapped.

To mitigate the second challenge \textbf{C2}, our approach involves initially employing noise suppression to enhance the signal-to-noise ratio. Subsequently, transforming the temporal imprints, inspired by their similarity to time-series signals, into the spectrum domain is undertaken to amplify the discernible differences (as detailed in \autoref{sec:synthetic_loss}).

To address the third challenge \textbf{C3} to ensure the efficacy against all forms of AE attacks, it is crucial \textit{not} to assume any prior knowledge of these attacks. We resolve this by framing adversarial sample detection as an anomaly detection problem. Leveraging a deep support vector data description (deep SVDD) trained solely on imprints derived from benign examples as a single-class classifier (as elaborated in \autoref{sec:overview}), we tackle this challenge head-on.

\noindent{\bf Contribution:} Our main contributions are threefold\footnote{Source code will be released upon publication of this work.}:

\noindent$\bullet$ To the best of our knowledge, \name pioneers the exploration of traceable temporal imprints to characterize adversarial behaviors, a significant change from existing methods that focus solely on spatial imprints. This novel paradigm not only overcomes several limitations of existing methods but also keeps the integrity of the protected model and its training procedure intact.

\noindent$\bullet$  By integrating temporal imprints into a universally adaptable loss metric and enhancing the differentiation between adversarial and benign examples through innovative techniques, we introduce the \name framework that operates effectively without prior AE knowledge. By design, \name is independent of data modalities, tasks, and model architectures, marking a significant advancement in adversarial defense.

\noindent$\bullet$  Comprehensive evaluations validate the exceptional performance of \name in detecting AEs with high accuracy and efficiency across different modalities, including images, audio, and text on up to 12 diverse attacks (including new audio AE attack of SMACK~\cite{yu2023smack}).
\name consistently achieves exceptional performance regardless of diverse datasets (four imagery including CIFAR10, STL10, CIFAR100 and TinyImageNet, two audio, and one textual classification datasets, and two regression datasets) and model architectures (e.g., ResNeXt50, BERT, RoBERT). For all cases, \name exhibits an AE detection accuracy exceeding 97\%, often around 99\%, maintaining a low false rejection rate of 1\%, underscoring its robustness and reliability. In addition, \name demonstrates high resilience against strong adaptive attacks.

\section{Insights}\label{sec:imprint}
This section begins with an overview of our key \textbf{insights and motivation}, highlighting the significance of universal temporal trajectory imprints \autoref{sec:imprint_intro} imprinted by AEs and their encapsulation within a universal loss metric \autoref{sec:synthetic_loss}. Inspired by these insights, we address two critical challenges of \textbf{C1} and \textbf{C2}, with the motivation of harnessing this universal metric to effectively identify AEs.

\subsection{New Signal: A Universal Temporal-Trajectory Imprint}
\label{sec:imprint_intro}

AEs, regardless of their generation methods, are designed to exhibit noticeable deviations from the normal data distribution, thus impacting inference results. Our exploration introduces a fresh perspective on identifying these deviations through trajectory imprints left across intermediate models (IMs) intrinsically rooted in the training procedure.

We specifically examine temporal-trajectory imprints captured by IMs at each training epoch. These imprints, serving as historical markers, reflect the model's evolution and only require preservation for analysis. Despite their similarities to the final deployed model, IMs offer unique insights with their epoch-specific parameter variations.

While various imprints like latent representations, logits, and softmax values are informative, they present challenges such as high dimensionality in scenarios with extensive class categories, or applicability limitations to common tasks like regression. In contrast, the loss metric stands out for its universal applicability and consistent lowest dimensionality across different tasks, making it an ideal candidate for a universal metric. This approach leverages the foundational role of loss functions in guiding the training process, addressing the challenge \textbf{C1} by adopting a loss-based imprint that transcends task and modality boundaries.

\subsection{Synthetic Loss: A Universal Temporal Imprint Metric}\label{sec:synthetic_loss}
A significant hurdle in this pipeline is the \textit{absence of ground truth} for incoming examples during inference, whether benign or adversarial, rendering the computation of their actual loss unfeasible. 
We note that works used trajectory for membership inference attacks~\cite{liu2022membership} or countering poisoning attacks~\cite{li2021anti} all require the available ground-truth as a must to fulfill their objectives. However, such a ground-truth is unavailable in our case. To circumvent this, we propose \textit{synthetic loss}, which diverges from the traditional loss computation that relies on known ground-truth annotations, such as labels. For tasks like classification or regression, this involves leveraging the output probabilities or predicted values from the target model as a `surrogate ground-truth', for each IM to compute the synthetic loss. Moreover, we have delved into additional strategies for synthesizing loss, which are elaborated upon in the \autoref{sec:losssysthesis}.

An example of synthetic loss imprinted trajectory across IMs in a classification scenario is demonstrated in~\autoref{fig:FGSM_fake_loss}.
This synthetic loss context mirrors the setup in \autoref{fig:FGSM_loss} that assumes the known ground-truth. Compared with \autoref{fig:FGSM_loss}, two critical observations emerge:

\observ{
The synthetic losses for adversarial and benign examples exhibit differences not only in their magnitude but also in the patterns of their fluctuations
}

\vspace{1pt}

\observ{
The clarity in distinguishing between adversarial and benign examples diminishes when employing synthetic loss, in contrast to the distinct separations with actual loss computations in \autoref{fig:FGSM_loss}
}

\noindent{$\blacksquare$} \textbf{Addressing C1}. 
The \textit{Observation 1}  underscores the potential of leveraging \textit{synthetic loss} as a universal metric for real-time identification of AE, without the need for additional training or alterations to the target model. Crucially, this approach bypasses the requirement for ground truth in loss computation, effectively overcoming challenge \textbf{C1}.

\noindent{$\blacksquare$} \textbf{Addressing C2}. 
The \textit{Observation 2}, however, poses a new challenge \textbf{C2}: trajectory imprints between benign and adversarial examples can be (significantly) overlapped, resulting in low distinguishability. 
We note that this imprint overlapping can be severe for some AE attacks or dependent on tasks e.g., regression. We have shown the synthetic loss imprint of a regression task (multivariate time-series forecasting) across IMs in~\autoref{fig:FGSM_overlap} (right)---experimental settings are detailed in \autoref{sec:setup}. We have also additionally shown the synthetic loss imprint of STL10 in~\autoref{fig:FGSM_overlap} (left), where the overlapping is extremely severe.

Nonetheless, the imprints left by AEs do differ from those of benign instances. 
IMs corresponding to early epochs still demonstrate relatively high randomness---different trends for adversarial and benign examples---without exhibiting substantial convergence. \textit{In essence, AEs, not originating from the same distribution as clean examples, manifest discernible statistical temporal-trajectory imprints}. Consequently, the challenge \textbf{C2} persists but may be effectively addressed by amplifying the temporal synthetic loss distinguishability. Two primary techniques are proposed to tackle challenge \textbf{C2}:

\noindent\(\bullet\) \textit{Noise Suppression}: 
The raw synthetic loss imprint contains noise, requiring salient feature extraction to enhance the signal-to-noise ratio. Therefore, we employ an LSTM auto-encoder for noise reduction, focusing on extracting critical features to improve the signal-to-noise ratio (SNR).

\noindent\(\bullet\) \textit{Spectrum Transformation}: In signal processing, transforming a time-series signal from the time-domain to the frequency/spectrum domain often reveals more distinct features. Given that loss imprints inherently resemble time-series data, applying spectrum transformation aligns naturally with their temporal characteristics, offering a promising avenue for addressing the challenges in AE detection.
Accordingly, we constructively transform the dimensionality-reduced synthetic loss feature into a spectrum to further accentuate the distinguishability between adversarial and benign examples. Equally, it further enhances the SNR.

\begin{figure}[ht]
    \centering
    \includegraphics[trim=0 0 0 0,clip,width=1.0\linewidth]{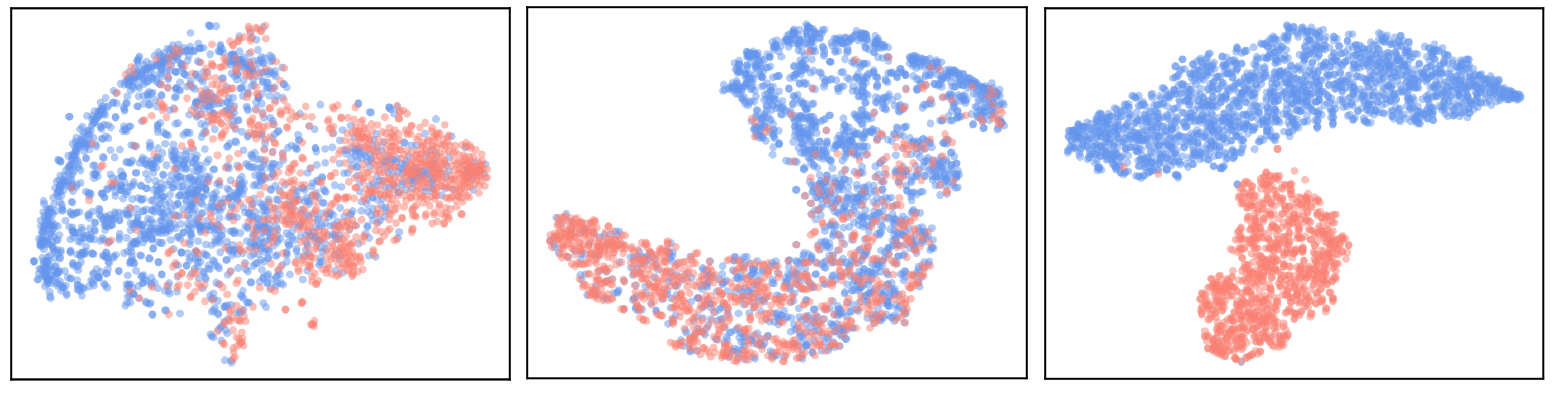}
        \caption{T-SNE visualization of benign and adversarial examples of (left) synthetic loss, (middle) after noise suppression, and (right) after spectrum transformation. Benign examples are with cornflowerblue color and adversarial examples are with salmon color. \textit{Quantitative evaluations} of the role of noise reduction and spectrum transformation are detailed in \autoref{sec:ablation}.
        }
    \label{fig:t-SNE_vis}
\end{figure}

\vspace{0.0cm}
\noindent\textbf{Visualization and Qualitative Analysis.} 
We illustrate the effect of noise suppression and spectrum transformation on the setting presented in \autoref{fig:FGSM_fake_loss}. Leveraging t-SNE~\cite{van2008visualizing}, we project both adversarial and benign examples onto a two-dimensional space, as depicted in \autoref{fig:t-SNE_vis}. Notably, the synthetic losses showcased in \autoref{fig:t-SNE_vis} (left) demonstrate a considerable overlap between adversarial and benign examples. In addition, the two clusters are not apparent.

Following noise suppression, executed through an LSTM-encoder outlined in \autoref{sec:implementation} for salient feature embedding, the overlap diminishes noticeably. Consequently, the two clusters become denser, as visualized in \autoref{fig:t-SNE_vis} (middle). Subsequently, upon transformation into the spectrum domain utilizing Fast Fourier Transform, a clear separation emerges between adversarial and benign examples, as demonstrated in \autoref{fig:t-SNE_vis} (right). It's important to note that in this spectrum depiction, only the amplitude information is displayed, while the phase information remains omitted.

\section{\name}\label{sec:design}
Having navigated through challenges \textbf{C1} and \textbf{C2}, we uncover a viable path for utilizing synthetic loss as a means to identify AEs post-deployment of the target model. This motivates us to develop \name framework. We begin by defining the threat model and addressing the last remaining challenge, \textbf{C3}. Subsequently, we overview and detail the \name.

\subsection{Threat Model}
\noindent\textbf{Attacker:} 
We consider a strong attacker, who is endowed with in-depth, white-box knowledge of the target model. This knowledge allows for the execution of advanced white-box AE attacks. Despite this profound insight, the attacker is barred from accessing the IMs (relaxed in adaptive attacks in Section~\ref{sec:adaptive}), which, though not as potent as the final model, contain valuable temporal data reflective of the model's training trajectory. The attacker is unconstrained by computational limits or the extent of perturbation applied to craft AEs, with a general preference for subtler modifications to ensure the deceptive alterations remain undetectable, thereby preserving the semantic integrity of the adversarial inputs.

\noindent\textbf{Defender:} 
The defender is typically the entity deploying the model. Their arsenal is limited to the deployed model, referred to as the target model, and its intermediate states, without the liberty to alter its training dynamics or access insights into the methodologies used by attackers for crafting AEs. This setup places the defender in a challenging position, armed with minimal knowledge, primarily relying on the IMs naturally generated at each training epoch and then preserved. These IMs, while not as refined as the final model, serve as a crucial resource for the defender and are preserved for potential defensive utility.

\begin{figure*}[h]
        
	\centering
	\includegraphics[trim=0 0 0 0,clip,width=0.80 \textwidth]{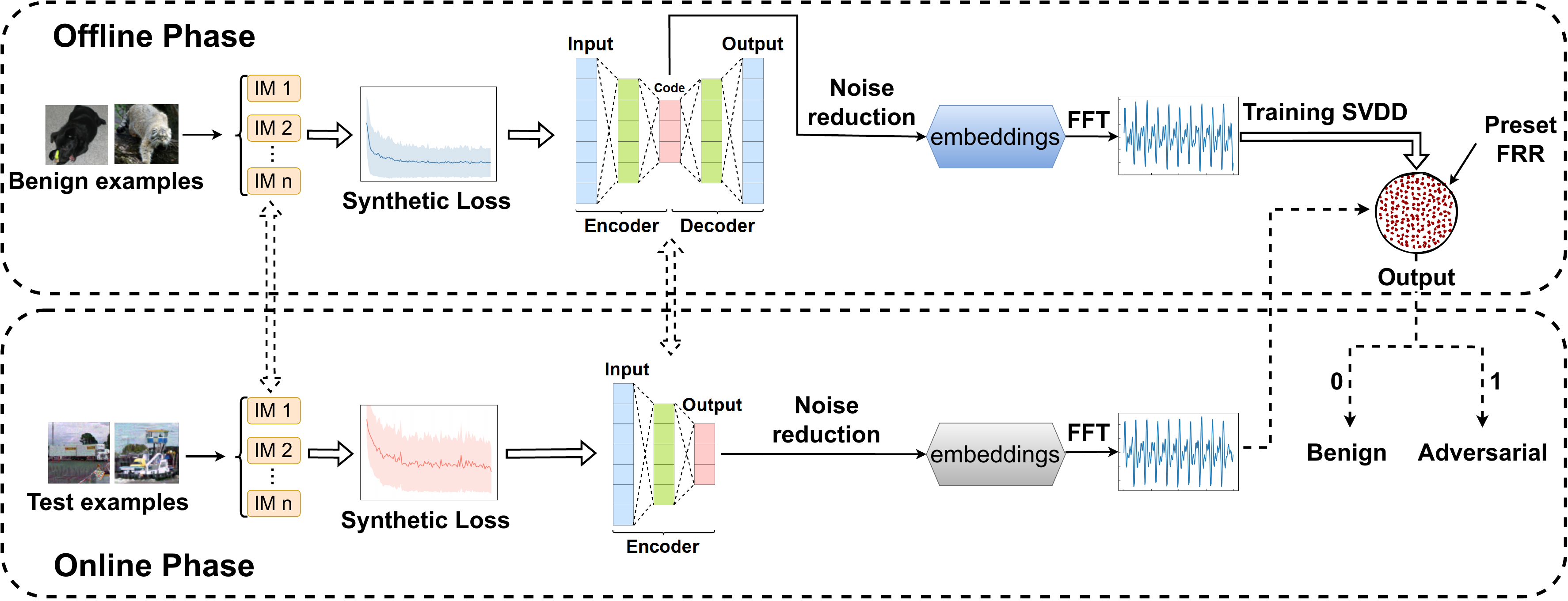}
	\caption{The \name overview. The IM stands for an intermediate model of the underlying/victim model.}
	\label{fig:TRAIT_overview}
     
\end{figure*}

\subsection{Overview of \name}\label{sec:overview}

\noindent{$\blacksquare$} 
\textbf{Addressing C3}. The absence of prior knowledge of AE crafting methods constitutes the third challenge, \textbf{C3}. This situation leaves the defender reliant solely on the analysis of benign examples' temporal trajectory imprints. To overcome \textbf{C3}, we integrate anomaly detection, specifically employing the SVDD~\cite{tax2004support,ruff2018deep} model. The SVDD is meticulously trained on the refined and spectrally transformed imprints of benign examples, ensuring a focused and effective detection mechanism. By addressing \textbf{C3} in conjunction with the previously circumvented challenges \textbf{C1} and \textbf{C2}, we lay a solid foundation for the development of \name, effectively equipping it to identify and mitigate adversarial threats without any prior knowledge of AE generation techniques. 

The overview of \name is depicted in \autoref{fig:TRAIT_overview}, consisting of two phases: the offline phase and the online phase. The former phase establishes an anomaly detector that identifies the AEs as anomalies throughout the subsequent online phase.

\subsubsection{Offline Phase}
During the offline phase, the model provider trains the DL model M with \(K\) epochs, the same as the typical training procedure. Each intermediate model, IM, per epoch is preserved by the provider. At the end of the training, there are \(K\) IMs, \{IM$_1$, \dots, IM$_k$, \dots IM$_K$\}. 
The IM$_k$ often serves as the final model M to be deployed in practice. Note that, the deployed model M can be the intermediate model generated at any epoch (usually close to the \(k_{\rm th}\) epoch), for example, with the highest validation accuracy. 

In step \circled{1}, the model provider randomly selects a subset of \(n\) validation examples (e.g., images), where \(n = e.g., 1000\), that are correctly classified as benign by the target model to their ground-truths. These examples are utilized as benign instances, excluding misclassified ones to mitigate noise. Each benign example undergoes evaluation by all \(K\) IMs. Subsequently, the synthetic loss for each \(k_{\text{th}}\) IM is calculated based on the softmax vectors in a classification task or the predicted value in a regression task, involving both the \(k_{\text{th}}\) IM and the target model \(M\). Typically, the target model M corresponds to the \(K_{\text{th}}\) IM. Consequently, for each benign example, \(K-1\) synthetic loss elements are generated.

In step \circled{2}, the synthetic loss imprints of the \(n\) benign examples are leveraged to train an encoder-decoder pair. Here, the bottleneck of this pair denotes the low-dimensional feature, termed the embedding, designed for noise reduction purposes. 
Subsequently, these embeddings undergo transformation into the spectrum domain and are utilized to train the SVDD. The SVDD functions as a single-class classifier, to detect AEs devoid of any prior knowledge regarding their attack methods. It's noteworthy that the preset false rejection rate—defined as the rate of falsely rejecting a benign example as an adversarial example—serves as a hyperparameter of the SVDD during its training phase.

\subsubsection{Online Phase} This phase is relatively straightforward. Given an incoming example, whether adversarial or benign, it is fed into all IMs to acquire the synthetic loss imprint trajectory (e.g., \(K-1\) elements). This loss trajectory is embedded by the encoder, and the resulting embedding is transformed into the spectrum domain before being fed into the SVDD that judges whether the input is benign or adversarial.

\subsection{Key Component Implementation}\label{sec:implementation}

The second step \(\circled{2}\) during the offline phase holds significant importance, encompassing two pivotal components: the discriminability intensifier and the deep-SVDD. The discriminability intensifier is comprised of an autoencoder and the spectrum transformation, which aims to enhance the discriminability between AE synthetic loss trajectory-imprints and benign counterparts. On the other hand, the deep-SVDD serves as an anomaly detector to identify AEs. 

\noindent\textbf{Discriminability Intensifier} consists of two parts: an encoder for low-dimensional synthetic loss embedding operation and a spectrum transformation operator. For the former, The use of an encoder to reduce dimensionality and extract salient features is primarily aimed at capturing essential information while discarding noise or irrelevant details in the synthetic loss imprint. 
Training an encoder-decoder pair facilitates this process by compressing the synthetic loss imprint into a lower-dimensional space (embedding) using the encoder, focusing on the most salient feature of the input. 
As the synthetic loss imprint is akin to a time-series signal, we thus leverage an LSTM-based encoder to perform the embedding. More specifically, we use a 2-layer bi-directionally structured LSTM model as the autoencoder for synthetic loss imprint noise and dimension reduction.
In order to prevent the model from overfitting, a dropout value of 0.2 was applied to both the encoder and decoder.
The number of training epochs is 150. Note the autoencoder training is fast. Also note that the decoder is no longer needed in the online phase of \name, shown in \autoref{fig:TRAIT_overview}. Because now only the encoder is required for the embedding operation. The latter spectrum transformation is through the FFT. Inspired by the fact that time-series signal analysis can be more efficient in the spectrum domain and the temporal imprints can indeed be treated as time-series signals along the epoch-axis. 

\noindent\textbf{Deep-SVDD} ~\cite{ruff2018deep} is a variant of the classical SVDD algorithm that extends its application to high-dimensional data by utilizing deep neural networks. Deep SVDD aims to learn a representation of normal (non-anomalous) data instances in a lower-dimensional space, often called the embedding space or hypersphere, while simultaneously minimizing the volume of this space. By doing so, it captures the most essential characteristics of the normal data, allowing anomalies to be detected as instances that fall outside this learned normality region in the embedding space.

The Deep-SVDD incorporates a preset FRR as one of its input hyperparameters\footnote{The Deep-SVDD is adopted from~\url{https://github.com/yzhao062/pyod/blob/master/examples/deepsvdd_example.py}. The SVDD function 
parameter \texttt{contamination} represents the preset percentage of benign samples being tolerated to be outliers, which is the preset FRR.}. The defender has the flexibility to set a small FRR (for example, 1-3\% in practical scenarios) during the Deep-SVDD training. This FRR signifies the acceptable probability that a benign example might be incorrectly classified as an AE, resulting in rejection. It's important to note that the trained Deep-SVDD operates as a binary classifier, making decisions based on its learned threshold to distinguish between adversarial and benign examples.

\section{Evaluation}\label{sec:evaluation}
This section extensively evaluates \name with image classification tasks. Its cross-modality and cross-task capability are affirmed in Section~\ref{sec:crossdomain} and Section~\ref{sec:crosstask}, respectively.

\subsection{Setup}\label{sec:setup}
\noindent\textbf{Dataset.} While two widely used benchmark datasets of CIFAR10~\cite{krizhevsky2009learning} and STL10~\cite{coates2011analysis} are extensively evaluated herein, we have evaluated CIFAR100 (see \autoref{sec:categoryNum}) and TinyImageNet (see \autoref{sec:complicatedDataset}). STL10 has 13,000 color images of size \(96 \times 96 \times 3\). The training and test sets have 5,000 and 8,000 images, respectively, with 10 classes in total.
The CIFAR10 dataset consists of 60,000 color images of size \(32 \times 32 \times 3\) in 10 classes, with 6$,$000 images per class. There are 50,000 training images and 10,000 test images.

\noindent\textbf{Attack.} We evaluate \name up to 7 mainstream AE attacks: 6 white-box and 1 black-box. Specifically, the white-box adversarial examples include Fast Gradient Sign Method (FGSM)~\cite{goodfellow2014explaining}, Projected Gradient Descent (PGD)~\cite{madry2017towards}, Basic Iterative
Method (BIM)~\cite{kurakin2018adversarial}, Carlini and
Wagner Attack (CW)~\cite{carlini2017towards}, DeepFool~\cite{moosavi2016deepfool}, Jacobian Saliency Map Attack (JSMA)~\cite{papernot2016limitations}. The black-box adversarial example attack is Boundary attack~\cite{brendel2017decision}. Each attack is detailed in Appendix~A. 

All these AE attacks use \(\ell \)-norm (e.g., \(\ell_0 \), \(\ell_1 \), \(\ell_2 \) and \(\ell_\infty \)) as perturbation constraints or measurements~\cite{machado2021adversarial}. 
In our evaluations, the \(\ell_2 \)-norm is set for DeepFool, \(\ell_0 \)-norm is set for JSMA, and \(\ell_\infty \)-norm is set for FGSM, BIM, PGD, CW, and Boundary attacks. Note that the perturbation magnitude \(\epsilon\) can be flexibly configured for FGSM, BIM, and PGD. 
While for other AE attacks, the \(\epsilon\) is usually fixed.

\noindent\textbf{Model.} We use ResNet18~\cite{he2016deep} for extensive evaluations in this section, while we have also taken VGG16~\cite{simonyan2014very} for generalization evaluations. 

\noindent\textbf{Machine.} The experiments were implemented on the DL framework PyTorch 1.13.1 and Python 3.9. The machine is a LENOVO laptop with an AMD Ryzen 7 5800H CPU (8 logical cores) and 16GB DRAM memory, and a GeForce GTX 1650 GPU (4GB video memory).

\noindent\textbf{Metric.} The metrics of detection accuracy and false rejection rate (FRR) are pivotal for assessing the effectiveness of \name's performance. Detection accuracy quantifies the likelihood of correctly identifying adversarial examples (AEs) as such. Conversely, the FRR gauges the likelihood of erroneously categorizing benign examples as AEs, leading to their rejection. Ideally, detection accuracy should approach 100\%, while the FRR should be minimized to 0\%.

\begin{table}[ht]
\centering 
\caption{\name detection performance (CIFAR10+STL10).
}
\resizebox{0.9\linewidth}{!}
{
\begin{tabular}{cccccc}
\toprule
\multirow{3}{*}{\begin{tabular}[c]{@{}c@{}}AE attack\\ Method\end{tabular}} &
  \multirow{3}{*}{\begin{tabular}[c]{@{}c@{}}Purturb \\ Magnitude\end{tabular}} &
  \multirow{3}{*}{\begin{tabular}[c]{@{}c@{}}Preset\\ FRR(\%)\end{tabular}} &
  \multicolumn{3}{c}{AE Acc(\%)} \\ \cmidrule(l){4-6} 
 &
   &
   &
  \multicolumn{2}{c}{CIFAR10} &
  STL10 \\ \cmidrule(l){4-5} \cmidrule(l){6-6}
 &
   &
   &
  \multicolumn{1}{c}{ResNet18} &
  \multicolumn{1}{c}{VGG16} &
  ResNet18 \\ \midrule
\multirow{6}{*}{\begin{tabular}[c]{@{}c@{}}BIM\\ (\(\ell_\infty \)-norm)\end{tabular}} &
  \multirow{3}{*}{$\epsilon = 8/255$} &
  1 &
  \multicolumn{1}{c}{97.94} &
  \multicolumn{1}{c}{98.61} &
  97.88 \\ 
 &
   &
  3 &
  \multicolumn{1}{c}{98.66} &
  \multicolumn{1}{c}{98.89} &
  98.12 \\ 
 &
   &
  5 &
  \multicolumn{1}{c}{99.11} &
  \multicolumn{1}{c}{99.37} &
  98.62 \\ \cmidrule(l){2-6} 
 &
  \multirow{3}{*}{$\epsilon = 16/255$} &
  1 &
  \multicolumn{1}{c}{97.95} &
  \multicolumn{1}{c}{98.43} &
  97.21 \\ 
 &
   &
  3 &
  \multicolumn{1}{c}{98.32} &
  \multicolumn{1}{c}{98.81} &
  97.86 \\ 
 &
   &
  5 &
  \multicolumn{1}{c}{98.84} &
  \multicolumn{1}{c}{98.95} &
  98.44 \\ \midrule

\multirow{6}{*}{\begin{tabular}[c]{@{}c@{}}FGSM\\ (\(\ell_\infty \)-norm)\end{tabular}} &
  \multirow{3}{*}{$\epsilon = 8/255$} &
  1 &
  \multicolumn{1}{c}{98.31} &
  \multicolumn{1}{c}{99.01} &
  98.16 \\ 
 &
   &
  3 &
  \multicolumn{1}{c}{98.80} &
  \multicolumn{1}{c}{99.33} &
  98.33 \\ 
 &
   &
  5 &
  \multicolumn{1}{c}{99.38} &
  \multicolumn{1}{c}{99.41} &
  99.11 \\ \cmidrule(l){2-6} 
 &
  \multirow{3}{*}{$\epsilon = 16/255$} &
  1 &
  \multicolumn{1}{c}{99.75} &
  \multicolumn{1}{c}{99.35} &
  99.23 \\ 
 &
   &
  3 &
  \multicolumn{1}{c}{99.76} &
  \multicolumn{1}{c}{99.72} &
  99.41 \\ 
 &
   &
  5 &
  \multicolumn{1}{c}{99.89} &
  \multicolumn{1}{c}{99.84} &
  99.44 \\ \midrule
\multirow{6}{*}{\begin{tabular}[c]{@{}c@{}}PGD\\ (\(\ell_\infty \)-norm)\end{tabular}} &
  \multirow{3}{*}{$\epsilon = 8/255$} &
  1 &
  \multicolumn{1}{c}{97.45} &
  \multicolumn{1}{c}{98.79} &
  97.46 \\ 
 &
   &
  3 &
  \multicolumn{1}{c}{98.01} &
  \multicolumn{1}{c}{99.11} &
  98.82 \\ 
 &
   &
  5 &
  \multicolumn{1}{c}{98.33} &
  \multicolumn{1}{c}{99.24} &
  98.91 \\ \cmidrule(l){2-6} 
 &
  \multirow{3}{*}{$\epsilon = 16/255$} &
  1 &
  \multicolumn{1}{c}{96.44} &
  \multicolumn{1}{c}{98.28} &
  96.26 \\ 
 &
   &
  3 &
  \multicolumn{1}{c}{96.89} &
  \multicolumn{1}{c}{98.56} &
  96.45 \\ 
 &
   &
  5 &
  \multicolumn{1}{c}{97.22} &
  \multicolumn{1}{c}{98.76} &
  97.03 \\ \midrule
\multirow{3}{*}{\begin{tabular}[c]{@{}c@{}}CW\\ (\(\ell_\infty \)-norm)\end{tabular}} &
  \multirow{3}{*}{-} &
  1 &
  \multicolumn{1}{c}{99.76} &
  \multicolumn{1}{c}{99.64} &
  99.42 \\ 
 &
   &
  3 &
  \multicolumn{1}{c}{99.76} &
  \multicolumn{1}{c}{99.81} &
  99.59 \\ 
 &
   &
  5 &
  \multicolumn{1}{c}{99.83} &
  \multicolumn{1}{c}{99.86} &
  99.71 \\ \midrule
\multirow{3}{*}{\begin{tabular}[c]{@{}c@{}}DeepFool\\ (\(\ell_2 \)-norm)\end{tabular}} &
  \multirow{3}{*}{-} &
  1 &
  \multicolumn{1}{c}{99.59} &
  \multicolumn{1}{c}{99.31} &
  99.46 \\ 
 &
   &
  3 &
  \multicolumn{1}{c}{99.70} &
  \multicolumn{1}{c}{99.44} &
  99.53 \\ 
 &
   &
  5 &
  \multicolumn{1}{c}{99.71} &
  \multicolumn{1}{c}{99.72} &
  99.64 \\ \midrule
\multirow{3}{*}{\begin{tabular}[c]{@{}c@{}}JSMA\\ (\(\ell_0 \)-norm)\end{tabular}} &
  \multirow{3}{*}{-} &
  1 &
  \multicolumn{1}{c}{99.56} &
  \multicolumn{1}{c}{99.64} &
  99.62 \\ 
 &
   &
  3 &
  \multicolumn{1}{c}{99.67} &
  \multicolumn{1}{c}{99.68} &
  99.78 \\ 
 &
   &
  5 &
  \multicolumn{1}{c}{99.78} &
  \multicolumn{1}{c}{99.77} &
  99.96 \\ \midrule
\multirow{3}{*}{\begin{tabular}[c]{@{}c@{}}Boundary\\ (\(\ell_\infty \)-norm)\end{tabular}} &
  \multirow{3}{*}{-} &
  1 &
  \multicolumn{1}{c}{99.73} &
  \multicolumn{1}{c}{99.15} &
  98.09 \\ 
 &
   &
  3 &
  \multicolumn{1}{c}{99.89} &
  \multicolumn{1}{c}{99.62} &
  98.43 \\ 
 &
   &
  5 &
  \multicolumn{1}{c}{99.97} &
  \multicolumn{1}{c}{99.78} &
  98.77 \\ \bottomrule
\end{tabular}
}
\label{tab:cifar10-results}
\end{table}

\subsection{Results}\label{sec:imageresults}
We first show and interpret the results of CIFAR10 using ResNet18 model and then the results of STL10 and a different VGG16 model. All results are summarized in \autoref{tab:cifar10-results}.

\noindent\textbf{CIFAR10.}
We conducted ResNet18 training for 104 epochs. The final model from the last epoch is utilized as the target model, exhibiting a classification accuracy of 94.03\%.

\begin{figure}[ht]
    \centering
    \includegraphics[trim=0 0 0 0,clip,width=0.35\textwidth]{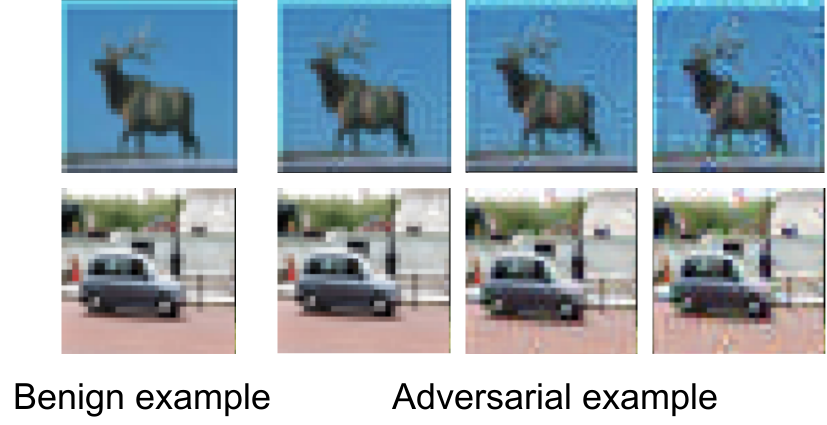}
        \caption{(Left) benign examples and (Right) their adversarial examples. From left to right, \(\epsilon\) = 8/255, 16/255 and 32/255, respectively. AE method is BIM.}
    \label{fig:bim_vis}
\end{figure}

We use the Adversarial Robustness Toolbox\footnote{\url{https://github.com/Trusted-AI/adversarial-robustness-toolbox}} to implement these AE attacks.
For all six white-box AE attacks, we use 3000 benign examples to generate adversarial examples per AE attack; the number of AEs depends on the attack success rate for each type of AE attack, e.g., varying from 77\% to 96\%. For configurable \(\epsilon\), we have set it to 8/255 and 16/255, respectively. These are preferable in practice by the attacker as the perturbation injected remains imperceptible---the perturbation appears to be perceptible when \(\epsilon=32/255\), as demonstrated in \autoref{fig:bim_vis}, \(\epsilon=32/255\) changed the image significantly compared to \(\epsilon=16/255\) and 8/255.

Considering the large amount of time/computation required to implement the black-box AE attack (in particular, the Boundary attack), we only select 200 samples in the test dataset (which takes about 10 hours) to generate AEs.

We have preset the FRR at 1\%, 3\%, and 5\%, respectively. As expected, there exists a trade-off where the detection accuracy tends to increase with higher FRR values. Remarkably, for CW, DeepFool, JSMA, and Boundary attacks, the detection accuracy is consistently near 100\%, even at an FRR as low as 1\%. It's important to note that the FRR is pre-set offline, aligning consistently with the online FRR measured using benign examples.
Regarding FGSM, the detection accuracy remains above 98\% across various perturbation magnitude configurations, even at an FRR of 1\%. For BIM and PGD attacks, the observed drop in detection accuracy is negligible (within 1\%) compared to FGSM.
Despite the stealthier nature of BIM and PGD attacks owing to their interactive perturbation injection, \name demonstrates efficient capture capabilities against these methods.

\noindent\textbf{STL10.}
We conducted training on ResNet18 using STL10 for 120 epochs, resulting in the last epoch model or target model achieving a classification accuracy of 76.89\%. 
To generate AEs for each white-box AE attack, we randomly selected 2000 benign samples. However, due to the significant computational and time overhead, we used only 200 benign examples (approximately 30 hours) to generate the AEs for the Boundary attack, the black-box attack. 
For the adjustable parameter \(\epsilon\), we set it to 8/255 and 16/255, respectively.

Even when considering a 1\% preset FRR, the detection accuracy remains consistently above 97\% against all AE attacks, as shown in the settings outlined in \autoref{tab:cifar10-results}. Notably, the detection performance on STL10 closely resembles that observed on the CIFAR10 dataset. This suggests that \name exhibits invariance across datasets.

\noindent\textbf{VGG16.} 
We conducted training on VGG16 using CIFAR10 for 105 epochs. The final model from the last epoch, which achieved a classification accuracy of 93.95\%, is deployed as the target model. The settings identical to those used to generate AEs for the ResNet18 model were employed to create AEs for the VGG16. As depicted in \autoref{tab:cifar10-results}, the performance of \name on VGG16 closely resembles that of ResNet18. This suggests that \name exhibits invariance not only across different dataset configurations but also across varying model architectures.

\section{Cross-Modality Tasks}\label{sec:crossdomain}
This section validates the inherent generic capability of \name across different modalities including audio and text.
\subsection{Audio}
\noindent\textbf{Setup.}
For the audio modality, the dataset of AudioMNIST~\cite{becker2018interpreting} is used. It consists of 30,000 audio samples of speech digits (0-9) from 60 different speakers. 
We implemented the task of recognizing 0-9 speech digits with AudioNet~\cite{becker2018interpreting}. Training and test datasets each have 25,000 and 5,000 samples.

\noindent\textbf{Results.}
We conducted training on AudioNet utilizing the AudioMNIST dataset for a total of 50 epochs. The final model obtained from the last epoch, which achieved a classification accuracy of 95.40\%, serves as the target model.

To generate AEs, we randomly selected 2000 benign audio samples. The implementation of adversarial attacks involved perturbing speech samples using six white-box AE methods: FGSM, PGD, BIM, CW, DeepFool, and JSMA. The AE attack generation toolbox is the same as for the image dataset. It's important to note that all these attacks are untargeted. Exemplified adversarial audio waveforms are depicted in Fig. 13 in the Appendix, which are almost similar to their benign ones. 


\begin{figure}[ht]
    \centering
    \includegraphics[trim=0 0 0 0,clip,width=0.45\textwidth]{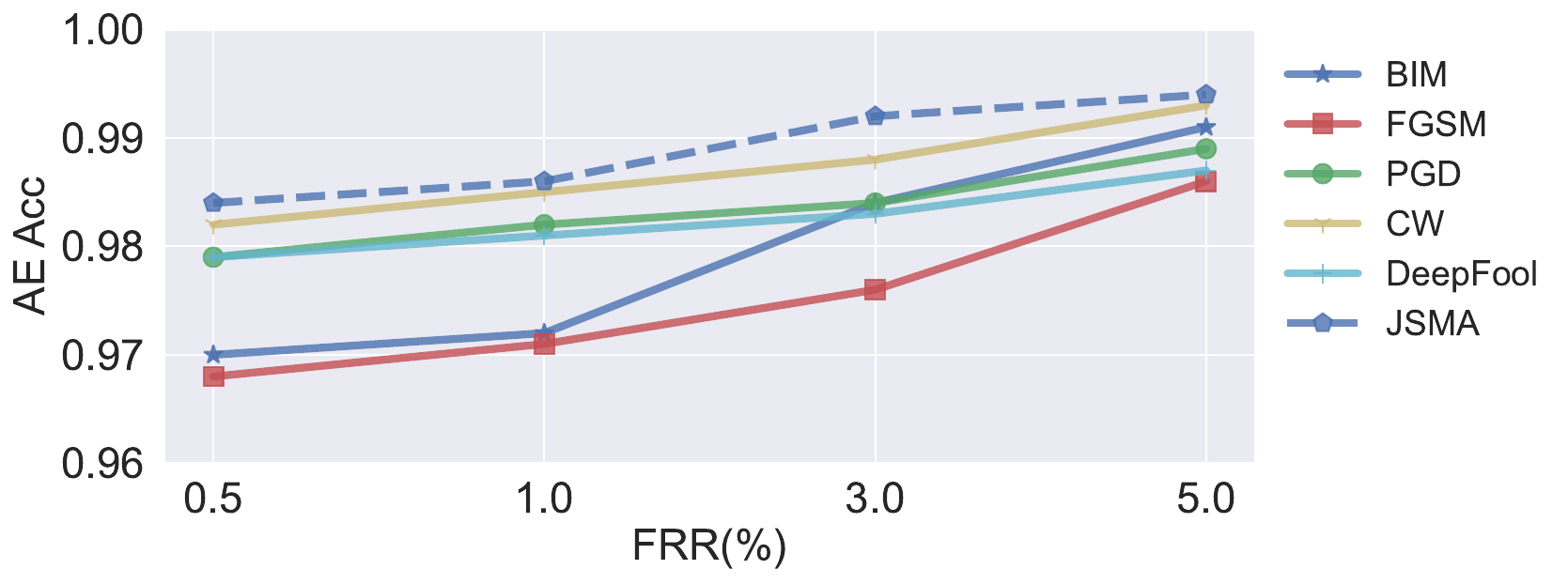}
        \caption{\name performance on AudioMNIST for six white-box AE attacks, where \(\epsilon\) = $5\times 10^{-5}$ for the three attacks of FGSM, BIM, and PGD that can be flexibly configured.}
    \label{fig:audio_results}
\end{figure}

For each of the six AE attacks, we assessed the efficacy of \name in detecting audio AEs by setting the FRR to predetermined levels of 0.5\%, 1\%, 3\%, and 5\%, respectively. The performance results are illustrated in \autoref{fig:audio_results}. 
It is evident that even with a preset FRR as low as 0.5\%, the detection accuracy of AE on the audio dataset consistently exceeds 97\% across all evaluated AE attacks. This affirms that beyond its performance in image-related tasks, \name demonstrates remarkable detection proficiency in a distinct audio modality.

\subsection{Textual}\label{sec:textual}
\noindent\textbf{Setup.}
The textual dataset employed in this study belongs to the Stanford sentiment treebank (SST)~\cite{socher2013recursive}, a sentiment analysis dataset publicly released by Stanford University. This dataset primarily revolves around sentiment categorization within movie reviews. SST-2, a subset of SST, focuses on binary categorization (positive or negative) and encompasses a total of 10,000 samples or sentences. 
Our experimentation involved training the SST-2 dataset using the BERT~\cite{devlin2018bert} model. 
We consider four mainstream textual AE attacks: Projected Word-Wise Substitution (PWWS)~\cite{ren2019generating}, TextBugger~\cite{li2018textbugger}, HotFlip~\cite{ebrahimi2018hotflip}, UAT~\cite{wallace2019universal}. Note that PWWS, HotFlip, and UAT are white-box attacks and TextBugger is a black-box attack. Each attack is detailed in Appendix~A. 


\noindent\textbf{Results.}
The BERT model was trained for 10 epochs on the SST-2 dataset, yielding a final epoch model deployed as the target model with a classification accuracy of 92.43\%. Additionally, the encoder embedding dimension is set to 6.

\begin{figure}[ht]
    \centering
    \includegraphics[trim=0 0 0 0,clip,width=0.35\textwidth]{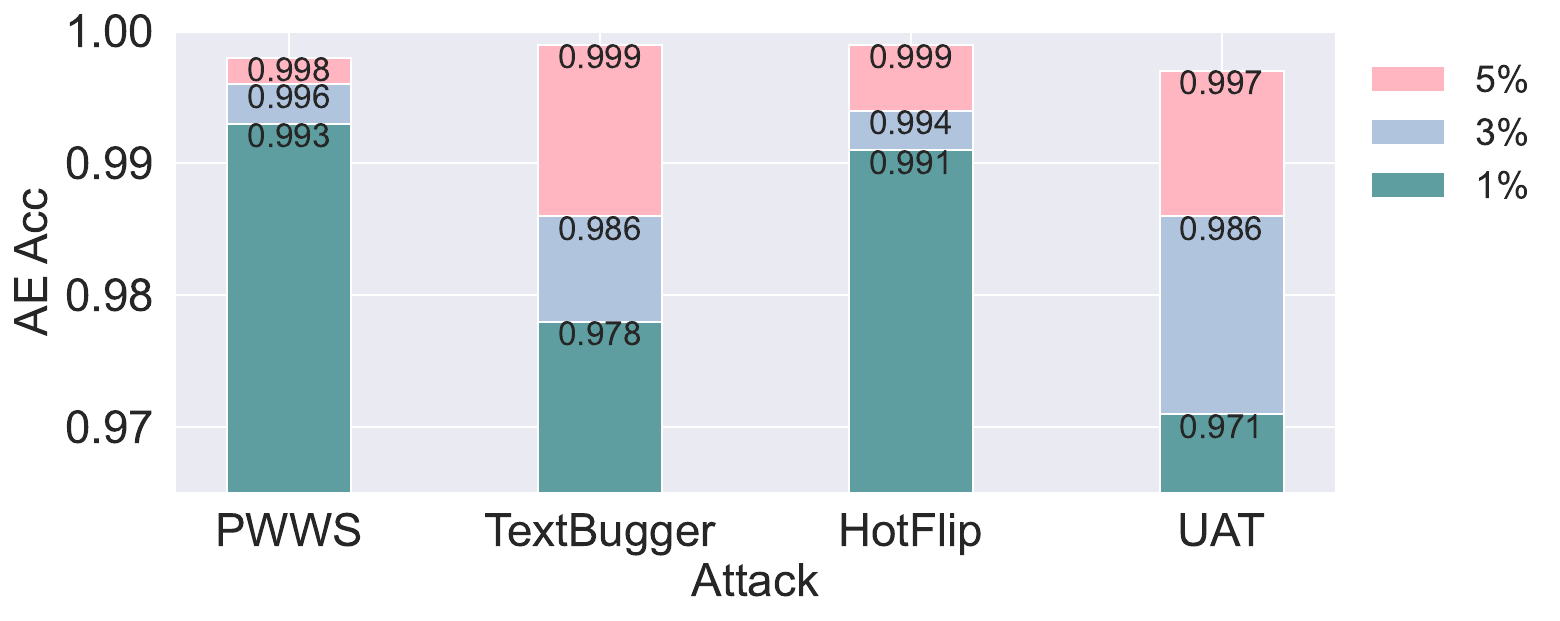}
        \caption{\name performance on SST-2 aganist four textual AE attacks. The target model is BERT.}
    \label{fig:text_results}
\end{figure}

These attacks were implemented using OpenAttack\footnote{\url{https://github.com/thunlp/OpenAttack}} by utilizing 1000 randomly chosen benign text samples.
As depicted in \autoref{fig:text_results}, we configured the FRR to 1\%, 3\%, and 5\% for each of these four textual adversarial attack methods. The results illustrate \name's exceptional detection performance against all four AE attacks, irrespective of white-box or black-box settings. Notably, the detection accuracy remains consistently above 97\% even at a minimal 1\% preset FRR. With a slightly higher tolerance of 3\% FRR, the detection accuracy surpasses 98.6\% across all AE attacks.
This validation underscores \name's versatility in the textual modality, demonstrating its high efficacy in identifying and capturing textual AEs.


\section{Non-Classification Tasks}\label{sec:crosstask}

\subsection{Time Series based Forecasting}

Time series forecasting is crucial in applications such as environmental parameters e.g., temperature, humidity, forecasting, stock market prediction, credit scoring, energy estimation, and traffic flow forecasting. 

\noindent\textbf{Setup.} In line with~\cite{chen2023targeted}, we utilized the New Zealand Land Temperatures dataset, a subset of Earth Surface Temperature Data 
along with a Bayesian LSTM model architecture. This dataset captures real-world surface temperature data. Similar to the approach in~\cite{chen2023targeted}, the dataset was partitioned into 1,459 training samples, 365 validation samples, and 100 test samples, each comprising 8 features---multivariate time-series. The input setup involved a three-step configuration for generating a one-step forecasting prediction. As for model architecture, we consider a 1-layer Bayesian LSTM, aligned with the architecture utilized in~\cite{chen2023targeted}.


\begin{table}[ht]
\centering 
\caption{\name detection performance of multivariate time-series temperature forecasting. }
\resizebox{0.60\linewidth}{!}
{
\begin{tabular}{ccc}
\toprule
\begin{tabular}[c]{@{}c@{}}AE Attack Method\end{tabular} & \begin{tabular}[c]{@{}c@{}}Preset FRR(\%)\end{tabular} & AE Acc(\%) \\ \midrule
\multirow{3}{*}{\begin{tabular}[c]{@{}c@{}}FGSM\\ ($\epsilon=0.2$)\end{tabular}} 
& 1 & 96.47 \\ 
& 3 & 96.85 \\  
& 5 & 98.23 \\ \midrule
\multirow{3}{*}{\begin{tabular}[c]{@{}c@{}}BIM\\ ($\epsilon=0.2$)\end{tabular}} 
& 1 & 95.46 \\ 
& 3 & 96.07 \\ 
& 5 & 96.17 \\ \bottomrule
\end{tabular}
}
\label{tab:time_series_results}
\end{table}

\noindent\textbf{Results.} We performed training for 500 epochs, and the model from the final epoch, which achieved a mean squared error of 0.0033, serves as the target model. Similar to~\cite{chen2023targeted}, our focus lies on targeted attacks, which hold more significance in forecasting tasks due to the attacker's ability to set desired targeted values flexibly.

The AE attacks---using the implementation of \cite{chen2023targeted}\footnote{\url{https://github.com/ProfiterolePuff/nvita}}---of BIM and FGSM are evaluated. As seen in \autoref{tab:time_series_results}, \name remains highly effective against the regression task, where a higher than 95\% detection accuracy is demonstrated even conditioned on a small FRR of 1\%. Note that non-classification tasks are often more challenging in the field of AE countering and much less explored compared to classification tasks. In addition, the results reaffirm that the \name is independent of the underlying model architecture, by noting the LSTM is a sequential model.


\subsection{Image based Age Estimation}
\noindent\textbf{Setup.}
We use the APPA-REAL face dataset that is for the regression task of estimating age~\cite{agustsson2017apparent}. APPA-REAL consists of 7,591 face images, where each image is annotated with real age and appearance age. Here we consider the real age when training the model. The dataset is split into 4,113 training images, 1,500 validation images and 1,978 test images. The image size is \(224 \times 224 \times 3\). The model architecture is ResNeXt50~\cite{xie2017aggregated}.
\begin{figure}[ht]
    \centering
    \includegraphics[trim=0 0 0 0,clip,width=0.4\textwidth]{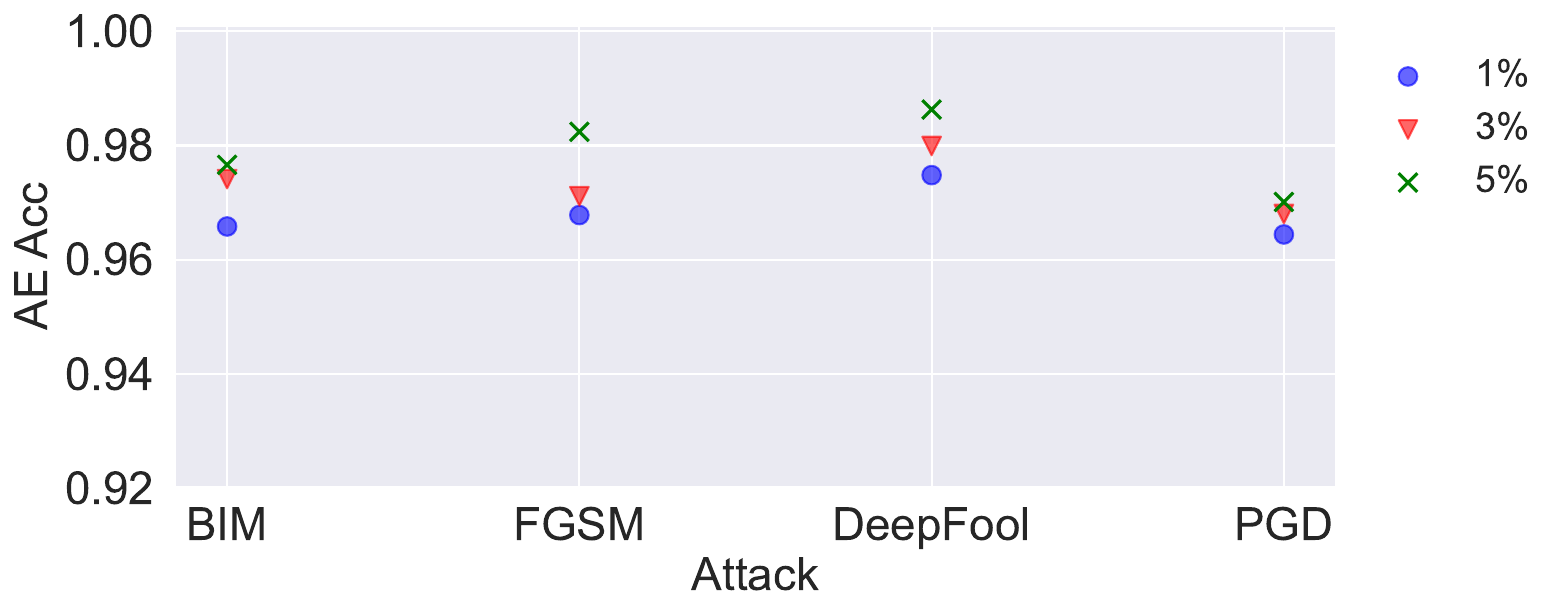}
        \caption{\name detection performance against image based age estimation.}
    \label{fig:img_regression}
\end{figure}

\noindent\textbf{Results.}
For the above time series regression, we have used targeted attacks. Now we consider the use of untargeted attacks. The age range is between 0.0 and 100.0. To ensure that the AE preserves the semantics of the benign example, the attack boundary is set within $\pm 4\%$, that is $\pm 4$ years of the predicted age given the benign image. We randomly selected 1,200 benign examples to create AEs with four attacks: FGSM, BIM, PGD, and DeepFool. We performed 60 epochs of training, and the model for the last epoch acting as the target model exhibits a mean square error of 4.76.

The results are shown in \autoref{fig:img_regression}. The detection accuracy is no less than 96\% even when the FRR is predetermined as small as 1\%. When a higher FRR is tolerable, the detection accuracy can be further improved. There are three implications. First, the \name is versatile to regression tasks other than classification. Second, the \name is universal to data types/modalities e.g., time-series data and image data. Third, the \name is effective in not only targeted but also non-targeted attacks under regression tasks.

\section{Adaptive Attack}\label{sec:adaptive}


There are many possible strategies for implementing different kinds of adaptive attacks. Exhausting them is out of the scope of this work, offering interesting future work. For our considered adaptive attack, we assume the attacker has the \textit{same access} to all IMs as the defender does, which is essentially out of the capability of an attacker in practice to evaluate the \textit{worst case} of \name under the strongest adaptive attack. In this context, the attacker adds a regularization constraint when crafting AEs to enforce that the loss trajectory of the AE through the course of IMs is the same (or very similar) as the trajectory of its benign example counterpart. Generally, the attacker tries to make the trajectory of AE and benign examples similar by setting a distance threshold (smaller distance, higher trajectory similarity). This objective is formulated as:

\begin{equation}\label{eq:adaptive}
\begin{aligned}
&\min_{x_{\rm adv}} \quad \mathcal{L}_{\rm adv}(x_{\rm adv}, y) + \lambda\cdot\textsf{Dist}_T(\textsf{T}(x_\text{adv}), \textsf{T}(x)) \\
&\text{s.t.} \quad \| x_{\rm adv} - x \|_p \leq \epsilon, 
\end{aligned}
\end{equation}

where \(\mathcal{L}_{\rm adv}(x_{\rm adv}, y)\) denotes the adversarial loss, which can be a classification loss, e.g., cross-entropy loss. \(\| x_{\rm adv} - x \|_p \leq \epsilon\) denotes the perturbation constraint between the AE \( x_{\rm adv} \) and the benign sample \( x \), where \( p \) is usually \(\infty\). Note this is a common constraint used in AE crafting to retain visual imperceptibility.

\textsf{T} denotes the extraction of synthetic loss values for all IMs, which can be expressed as: 
\begin{equation}\label{eq:T}
\begin{aligned}
&\textsf{T}(x) = [\mathcal{L}_1(x), \mathcal{L}_2(x), \dots, \mathcal{L}_{(K-1)}(x)],
\end{aligned}
\end{equation}

where \(\mathcal{L}_k(x)\) denotes the synthetic cross-entropy loss of the input sample \(x\) between the first \(k\) IM and the final model, i.e. \(\mathcal{L}_k(x) = \text{Loss}(f_k(x), f_K(x))\).
\(K\) denotes the number of all saved IMs---note that $K_{\rm th}$ IM is essentially the final model.
\(f_k(x)\) denotes the softmax output of the input sample \(x\) at the \(k_{\rm th}\) IM.

$\textsf{Dist}_T$ constrains the trajectory distance between the benign examples and its corresponding AE through e.g., L$_2$ norm, which is expressed as: 


\begin{equation}\label{eq:l2}
\begin{aligned}
\textsf{Dist}_T(\mathcal{L}(x'), \mathcal{L}(x)) = \frac{\sum_{k=1}^{K} (\mathcal{L}(x'_k) - \mathcal{L}(x_k))^2 - \text{Dist}_T^{\text{min}}}{\text{Dist}_T^{\text{max}} - \text{Dist}_T^{\text{min}}}.
\end{aligned}
\end{equation}

In this formula, \(\text{Dist}_T^{\text{max}}\) and \(\text{Dist}_T^{\text{min}}\)
denote the maximum and minimum values of the distance function \(\textsf{Dist}_T(\mathcal{L}(x'), \mathcal{L}(x)\), respectively, used for normalization.

In Eq.~\ref{eq:adaptive}, \(\lambda\) (1.0 is used) is the regularization factor that controls the weight of the AEs and benign examples of similarity terms in the total loss.
Eq.~\ref{eq:adaptive} sets two main goals that an attacker considers when generating AEs: misleading the final model to misclassify (the first loss term) and generating AEs with low trajectory distance to the benign example (the second L$_2$ regularising constraint term) to adaptively evade \name
 , while satisfying the perturbation constraint in Eq.~\ref{eq:adaptive}.

\begin{figure}[ht]
    \centering
    \includegraphics[trim=0 0 0 0,clip,width=0.50\textwidth]{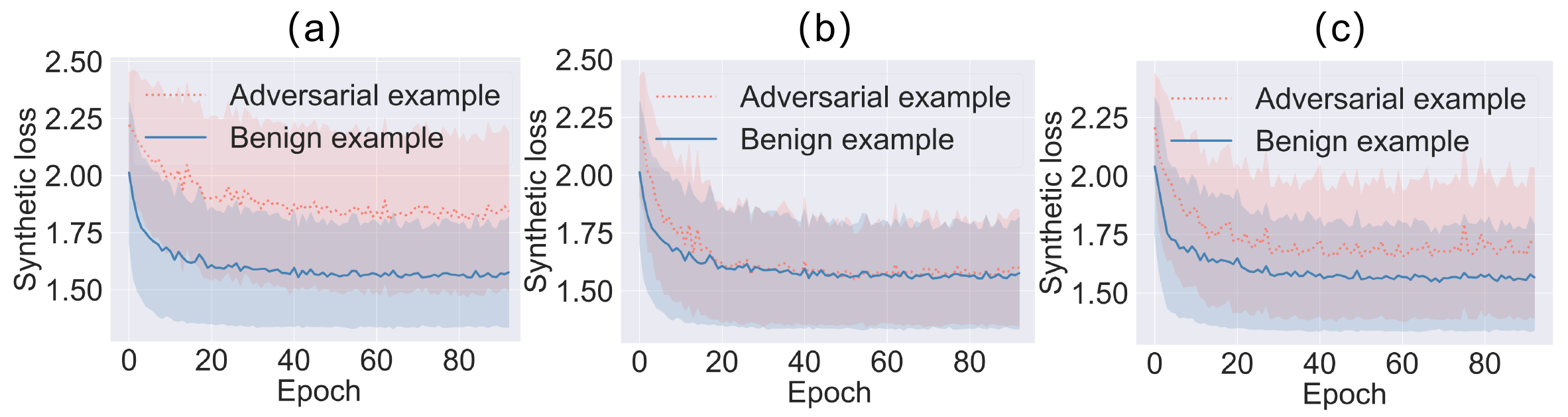}
        \caption{Comparison of trajectory overlap (a) without adaptive attack, and with adaptive by (b) using defender's IMs and (c) surrogate IMs.}
    \label{fig:adaptive_attack}
\end{figure}

By increasing the trajectory similarity, the attacker expects to evade the \name detection.  We use ResNet18+CIFAR10 and FGSM AE method for experiments.  On the one hand, by checking the trajectory of the AE and benign samples, as shown in~\autoref{fig:adaptive_attack} (b), they are indeed almost overlapped visually when greatly enforcing the similarity.  On the other hand, in the case shown in~\autoref{fig:adaptive_attack} (b), the success rate of crafting an AE substantially drops to only 3.3\% by setting the distance threshold to 0.19 
(in comparison, the success rate of AE is 76\% with a distance threshold of 0.52). 
Nonetheless, our \name still exhibits high detection accuracy of up to 95.20\% with an online FRR of 2.44\%. The reason is that the AE trajectory is still discernible (though small) from its benign example counterpart, especially in early epochs, which our innovative \name can still capture. 

As accessing the defender’s IMs is almost impossible in practice, we further consider a scenario where the attacker can train those IMs, namely surrogate IMs, by accessing the training dataset and knowledge of exact model/training parameters---such assumptions are essentially still strong as well. Then the attacker crafts the AEs following the procedure above but on the surrogate IMs. As shown in~\autoref{fig:adaptive_attack} (c), the AE trajectory becomes more discernable than those crafted upon the defender’s IMs, which is under expectation. For those AEs, by setting the same distance threshold of 0.19 when crafting AEs, the detection rate of \name now exceeds 99\% with an online 1.4\% FRR.

In summary, through the considered adaptive attacks, an attacker is able to craft AEs that share a similar trajectory with their benign counterparts but sacrifice a very low success rate. Even for those successful AEs, the \name still maintains a high detection performance. 

\section{Scalability}

We demonstrate the \name scalability regarding typical DL models (e.g., ResNet) and large models (e.g., pretrained natural language model). The former model is normally trained for a specific task. The latter first trains a foundation model and then fine-tunes it to downstream tasks.

\subsection{Typical DL Model}\label{sec:typicalDL}
The IMs need to be stored. Even though it is acceptable to take the trade-off between the memory and model security robustness, especially for security sensitive applications e.g., facial recognition and malware detection, it is still preferable to reduce the memory overhead to store these IMs. 

\begin{table}[ht]
\centering 
\caption{\name detection performance when only early epochs (i.e., the first 10, 20 and 30) synthetic losses are used (CIFAR10+ResNet18).}
\resizebox{0.80\linewidth}{!}
{
\begin{tabular}{cccccc}
\toprule
\multirow{2}{*}{\begin{tabular}[c]{@{}c@{}}AE Attack\\ Method\end{tabular}} &
  \multirow{2}{*}{\begin{tabular}[c]{@{}c@{}}Preset\\ FRR(\%)\end{tabular}} &
  \multicolumn{4}{c}{AE Acc(\%)} \\ \cmidrule(l){3-6} 
                      &   & \multicolumn{1}{c}{Epochs=10} & \multicolumn{1}{c}{Epochs=20} & \multicolumn{1}{c}{Epochs=30} & Epochs=104 \\ \midrule
\multirow{3}{*}{FGSM} & 1 & \multicolumn{1}{c}{95.88}     & \multicolumn{1}{c}{96.11}     & \multicolumn{1}{c}{96.42}     & 98.31      \\  
                      & 3 & \multicolumn{1}{c}{96.79}     & \multicolumn{1}{c}{97.32}     & \multicolumn{1}{c}{97.37}     & 98.80      \\  
                      & 5 & \multicolumn{1}{c}{97.17}     & \multicolumn{1}{c}{97.41}     & \multicolumn{1}{c}{97.55}     & 99.38      \\ \midrule
\multirow{3}{*}{PGD}  & 1 & \multicolumn{1}{c}{95.07}     & \multicolumn{1}{c}{95.18}     & \multicolumn{1}{c}{95.80}     & 97.45      \\ 
                      & 3 & \multicolumn{1}{c}{95.89}     & \multicolumn{1}{c}{96.04}     & \multicolumn{1}{c}{96.08}     & 98.01      \\  
                      & 5 & \multicolumn{1}{c}{96.93}     & \multicolumn{1}{c}{97.13}     & \multicolumn{1}{c}{97.36}     & 98.33      \\ \bottomrule
\end{tabular}
}
\label{tab:memory}
\end{table}

The disparity in trajectory imprint between benign and adversarial examples is notably prominent in the initial epochs, as depicted in Figures~\ref{fig:FGSM_fake_loss} and \ref{fig:FGSM_overlap}. So we may solely rely on these early epoch imprints to capture AEs, which could potentially be sufficient. This approach could substantially reduce stored IMs and, thus the memory overhead. 

To validate this hypothesis, we computed the synthetic loss using only the first 10, 20 and 30 IMs (each IM corresponds to one epoch), respectively, and the final model for the CIFAR10+ResNet18 combination. Specifically, PGD and FGSM were evaluated with $\epsilon=8/255$. The results are presented in \autoref{tab:memory}. Notably, when the number of IMs is limited to as small as 10, there is only a marginal drop in \name detection accuracy (e.g., less than 2.5\%) for both FGSM and PGD in comparison to the results in \autoref{tab:cifar10-results} (also copied into the last column where number of epochs is 104) at preset FRR values of 1\%, 3\%, 5\%, respectively. This reduction in memory size is approximately 90\%. Remarkably, the detection accuracy remains more than 95.0\% even at a minimal 1\% FRR.

In addition to reducing the number of IMs, further reductions in the model size of each IM can be achieved through various post-training DL model compression techniques, including quantization, weight clustering, and pruning. These techniques, capable of significantly reducing model size while preserving performance, are supported by leading commercial machine learning frameworks such as TensorFlow-Lite and PyTorch Mobile~\cite{ma2023quantization,TFLite,PytorchMobile}. 
Importantly, each of these compression operations can be separately applied through corresponding APIs, or \textit{combined for more comprehensive compression}. Furthermore, as consecutive IMs share similarities, weight sharing across consecutive IMs might be feasible. 

\subsection{Large Model}
As for large pretrained language models, they can have hundreds of millions of parameters (e.g., RoBERTa~\cite{liu2019roberta}, with a parameter count of about 120 million). These models are trained for general tasks and then can be customized to downstream tasks. It is absolutely unwise to fine-tune the entire model for downstream tasks, where parameter-efficient fine-tuning (PEFT) methods~\cite{han2024parameter} such as Low-Rank Adaptation (LoRA)~\cite{hu2021lora} are often leveraged by fine-tuning only a small fraction of (additional) model parameters to save computational overhead.

We use LoRA to fine-tune RoBERTa which is one of the large language models. The LoRA adds a small set of trainable parameters (665,858 parameters) while freezing the entire RoBERTa with 125,313,028 parameters. The fine-tuning parameters account for only 0.53\% of all parameters. We fine-tune 10 epochs using the SST-2~\cite{socher2013recursive} dataset for utterance sentiment classification, and the final CDA is 92.44\%. 
For PWWS~\cite{ren2019generating} and HotFlip~\cite{ebrahimi2018hotflip} attacks, the detection accuracy is 98.55\% and 98.24\% at preset FRR of 1\%, respectively. The additional storage memory overhead by incorporating \name is only about 5.3\% ($0.53\% \times 10$) considering 10 epochs. This fully demonstrates the scalability of \name on large models.

\section{Discussion}\label{sec:discussion}
\subsection{Component Ablation Study}\label{sec:ablation}
\name component’s functionality has been qualitatively justified in \autoref{sec:synthetic_loss} (particularly visualized in \autoref{fig:t-SNE_vis}), we now quantify core components functionality.

The CIFAR10 dataset with ResNet18 is used to assess the effectiveness of each component of \name. Specifically, we crafted 2,000 AEs using the BIM method ($\epsilon = 8/255$) and evaluated \name's performance under three different settings: i) removing noise reduction and FFT (used for spectrum transformation); ii) removing noise reduction; and iii) removing FFT. The performance of \name under each setting is detailed in~\autoref{tab:ablation}.

The results demonstrate that \name fails to achieve satisfactory performance if any component is omitted. With a preset 1\% FRR, removing the noise reduction component results in a decrease in detection accuracy from 97.94\% to 87.43\%, representing a drop of approximately 10\%. Removing the FFT or spectrum transformation causes an even more significant drop in detection accuracy, by 55\%. When both FFT and noise reduction are excluded, the detection accuracy plummets to just 9.32\%, rendering it useless.

These quantifications underscore the importance of \name's innovative component design, particularly the incorporation of spectrum transformation and noise reduction, with spectrum transformation being the most critical component.


\begin{table}[ht]
\centering 
\caption{\name's components ablation study. }
\resizebox{0.60\linewidth}{!}
{
\begin{tabular}{ccc}
\toprule
\begin{tabular}[c]{@{}c@{}}Removed Component\end{tabular} & \begin{tabular}[c]{@{}c@{}}Preset FRR(\%)\end{tabular} & AE Acc(\%) \\ \midrule
{\begin{tabular}[c]{@{}c@{}}Noise reduction\end{tabular}} 
& 1 & 87.43 \\ \midrule
\begin{tabular}[c]{@{}c@{}}FFT\\(Spectrum Transformation) \end{tabular}
& 1 & 43.20 \\ \bottomrule
\begin{tabular}[c]{@{}c@{}}Noise reduction + FFT \end{tabular}
& 1 & 9.32 \\ \bottomrule
\begin{tabular}[c]{@{}c@{}} \name \end{tabular}
& 1 & 97.94 \\ \bottomrule
\end{tabular}
}
\label{tab:ablation}
\end{table}

\subsection{More Attack}
To further demonstrate the \name generality to unknown attacks, we evaluate its performance on a recent AE attack on automatic speech recognition that is SMACK~\cite{yu2023smack} (USENIX 2023')---modifying inherent speech attribute of prosody. SMACK was shown to be evasive SOTA defenses~\cite{yu2023smack}. Following the SMACK setup, we evaluated \name. SMACK utilizes a generative model to produce adversarial examples. The victim model is DeepSpeech 2~\cite{amodei2016deep} trained on LibriSpeech~\cite{panayotov2015librispeech}.

We trained DeepSpeech 2 for 70 epochs and saved 70 IMs. The last epoch model is the deployed target model that has a word error rate of 10.51\% in the LibriSpeech dataset. 
Note that SMACK generates adversarial audio using the Common Voice~\cite{ardila2019common} dataset, which contains 81,085 human speech sentences, as the raw audio data. We randomly selected 500 examples from Common Voice, and the generated adversarial examples' attack success rate on DeepSpeech 2 is 87.6\%, similar to the SMACK reported attack success rate of 88.3\%. 
\textit{\name detection accuracy for these successful adversarial audio examples is 95.74\%, 96.21\%, 96.76\% at preset FRR of 1\%, 3\%, 5\% respectively}.

\subsection{Category Number}\label{sec:categoryNum}
Here, we evaluate the \name sensitivity to the number of categories in the classification tasks. The dataset of CIFAR100 with 100 classes is now used and the model of VGG16 is used. Different subsets of CIFAR100 with different numbers of categories are used for evaluations. The number is set to 20, 40, 60, and 80, respectively, and the whole dataset of 100 classes is also considered. We extensively evaluate the \name with FGSM and BIM attacks, both of which were evaluated at $\epsilon = 8/255$.

\begin{figure}[ht]
    \centering
    \includegraphics[trim=0 0 0 0,clip,width=0.50\textwidth]{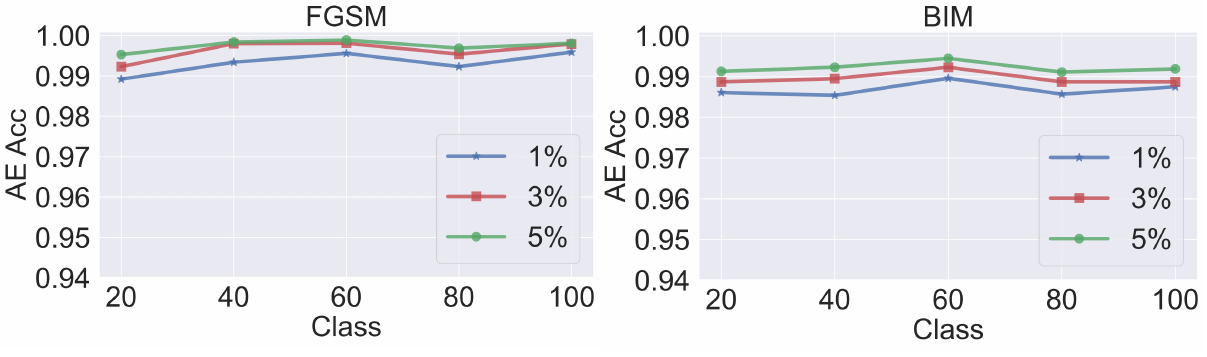}
        \caption{\name is independent of the number of classes. CIFAR100 dataset is used.}
    \label{fig:class}
\end{figure}
The results are depicted in \autoref{fig:class}. For each number of classes, we trained 130 epochs, all of which generated about 1500 adversarial examples. We preset FRR to 1\%, 3\% and 5\%, respectively. We find that changes in the number of classes do not cause notable changes in the detection capability of \name. In other words, \name is insensitive to changes in the number of classes. In addition, the AE detection accuracy is higher than 98\% even given a 1\% FRR, which is consistent with the results of CIFAR10 and STL10 in \autoref{tab:cifar10-results}.

\section{Latency Overhead}\label{sec:latency}
Compared to the standard inference latency of the target model, \name introduces additional sequential operations: loss synthesis, dimension reduction through an encoder, and assessment of adversarial behavior via Deep-SVDD. Importantly, the inference operation of IMs compared with the target model is not sequential, allowing for parallel execution. Thus, we quantify the time overhead for each sequential component using the VGG16 model and the CIFAR10 dataset, with 1000 examples dedicated to validation.

The average time overhead per image for target/IM model inference, loss synthesis, dimension reduction, and adversarial behavior detection through Deep SVDD is 0.47ms, 0.004ms, 0.06ms, and 0.07ms, respectively. Notably, all computations are conducted on a single CPU. Given that IMs operate in parallel with the target model, the additional latency introduced by \name including loss synthesis, dimension reduction, and adversarial behavior detection, is approximately 28.5\% (0.136ms to 0.47ms), which remains completely acceptable for enhanced security benefits.

\subsection{More Complicated Dataset}\label{sec:complicatedDataset}
For the image datasets, We have extensively evaluated CIFAR10 and STL10 for comprehensive validations in \autoref{sec:evaluation}. In \autoref{sec:categoryNum}, we have also evaluated CIFAR100. We now further evaluate on a more complicated dataset, TinyImageNet~\cite{le2015tiny}. We trained 200 epochs on ResNet18, with a 67.37\% CDA. We used only the first 30 IMs for \name, which is to further validate the scalability as in \autoref{sec:typicalDL}. \name detection performance against FGSM and PGD is shown in~\autoref{tab:tiny_results}. \name has a detection accuracy of not less than 96\%, which shows that \name still maintains an excellent performance on more complicated datasets, even when only using a small number of IMs. As TinyImageNet has 200 classes, this also validates the \name is insensitive to the number of categories.

\begin{table}[ht]
\centering 
\caption{\name detection performance on TinyImageNet. }
\resizebox{0.60\linewidth}{!}
{
\begin{tabular}{ccc}
\toprule
\begin{tabular}[c]{@{}c@{}}AE Attack Method\end{tabular} & \begin{tabular}[c]{@{}c@{}}Preset FRR(\%)\end{tabular} & AE Acc(\%) \\ \midrule
\multirow{3}{*}{\begin{tabular}[c]{@{}c@{}}FGSM\\ ($\epsilon=8/255$)\end{tabular}} 
& 1 & 96.60 \\ 
& 3 & 97.45 \\  
& 5 & 97.70 \\ \midrule
\multirow{3}{*}{\begin{tabular}[c]{@{}c@{}}BIM\\ ($\epsilon=8/255$)\end{tabular}} 
& 1 & 96.15 \\ 
& 3 & 96.54 \\ 
& 5 & 97.12 \\ \bottomrule
\end{tabular}
}
\label{tab:tiny_results}
\end{table}




\subsection{Imprint Types}\label{sec:imprint}
In addition to the universal metric of loss, we expect that the exploration of alternative imprints such as latent representation and softmax may exhibit more pronounced discrepancies between adversarial and benign examples. These imprints, capturing finer-grained information than the loss metric, offer a promising avenue for analysis. However, they often lack universality; for instance, while softmax is prevalent in classification tasks, its applicability to non-classification tasks remains limited. Moreover, these alternative imprints typically exhibit higher dimensionality compared to the loss. Consider a scenario involving facial recognition having 10,000 identities with 100 IMs: the dimensionality of the softmax imprint to the encoder amounts to $10,000 \times 100$, a significantly higher dimension than the $1 \times 100$ dimensionality of the loss imprint. Nevertheless, exploring alternative imprints holds merit, especially if they offer superior discriminative capabilities in \textit{specific} applications.

\begin{figure}[h]
    \centering
    \includegraphics[trim=0 0 0 0,clip,width=0.35\textwidth]{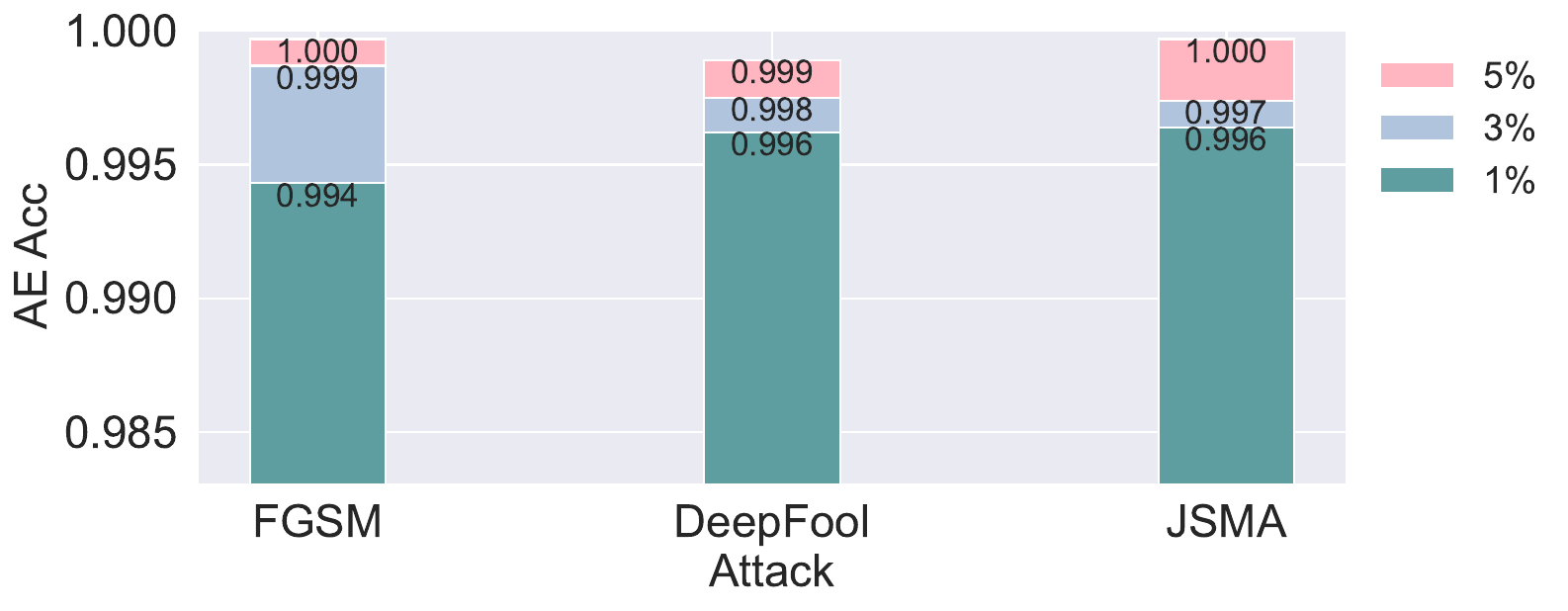}
        \caption{\name detection performance when the softmax is used as an imprint (CIFAR10+ResNet18).}
    \label{fig:softmax_results}
\end{figure}

\textit{Softmax Imprint:} To assess the efficacy of employing softmax as an imprint, we evaluated the performance of \name using CIFAR10 and ResNet18. Notably, CIFAR10 has ten categories, rendering the softmax imprint to the LSTM-autoencoder akin to a 10-channel/feature time-series signal. The evaluation settings align with those detailed in \autoref{sec:evaluation}. For adversarial attacks, we employed FGSM ($\epsilon=8/255$), DeepFool, and JSMA. The outcomes, illustrated in \autoref{fig:softmax_results}, demonstrate an enhanced detection accuracy while maintaining a consistent FRR. For instance, the detection accuracy against FGSM improved from 98.31\% (as indicated in \autoref{tab:cifar10-results}) to 99.40\% given a 1\% FRR. This improvement underscores the effectiveness of leveraging softmax directly, which encodes finer-grained information and better captures adversarial examples.

\subsection{Loss Synthesis}\label{sec:losssysthesis}

When synthesizing loss, all previous experiments use each of IMs softmax and the target model's softmax for computing the loss. Here we explore other loss synthesis means. Specifically, we synthesize the loss using the softmax of every two consecutive IMs.
The detection accuracy is summarized in Table 7 in Appendix, where three AE attacks are evaluated and all the experimental settings are the same as \autoref{sec:evaluation} (CIFAR10+VGG16). The performance using loss synthesized upon two consecutive IMs (second last column) is slightly lower than the performance using loss synthesized upon softmax of each IM and the target model (last column replicated from \autoref{tab:cifar10-results} to ease comparison). Nonetheless, the detection accuracy is still no less than 98\% even given a 1\% FRR.

\vspace{-0.2cm}
\subsection{Comparison}

This work mainly aims to \textit{uncover a new paradigm of understanding the temporal traces left by adversarial attacks}, which is later demonstrated to be highly effective in capturing such attacks.
Considering that \name does not necessitate prior knowledge of AE attacks, falling within the class of unsupervised AE detection, we conducted a comparative analysis of \name against two unsupervised AE detection methods. Specifically, they are NIC \cite{ma2019nic} (NDSS '2019), and ContraNet~\cite{yang2022you} (NDSS '2022), which we reproduced these two works.

\begin{figure}[ht]
	\centering
	\includegraphics[trim=0 0 0 0,clip,width=0.4\textwidth]{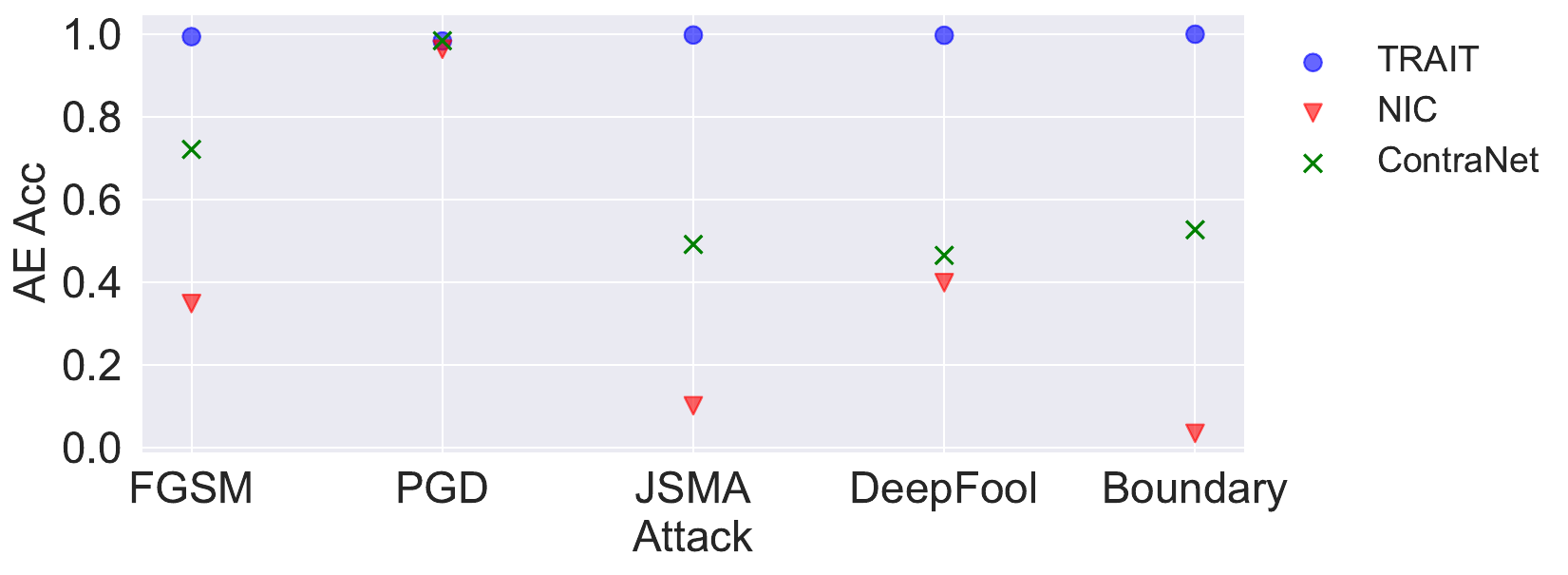}
	\caption{Comparison of \name with NIC and ContraNet.}
	\label{fig:comparison}
\end{figure}

We employed five attack methods, namely FGSM, PGD, JSMA, DeepFool and Boundary attack, for comparative validation. For FGSM, PGD, the $\epsilon$ was set to $8/255$, other experimental settings are the same as in~\autoref{tab:cifar10-results}. 
The comparison results are summarized in Figure~\ref{fig:comparison}, and the preset FRR was set to 5\%. This relatively high FRR of 5\%---\name still performs well given a 1\% FRR---is to align with settings of NIC and ContraNet, their detection accuracy is even worse when a lower FRR is set. 
It is worth mentioning that TRAIT, NIC and ContraNet have excellent performance in detecting PGD AEs with accuracies of 98.33\%, 96.33\% and 98.40\%, respectively. However, NIC and ContraNet are much lower than \name in detecting the other four AE methods.

\vspace{0.3cm}
\section{Related Work}\label{sec:relatedwork}

To counter AE attacks, three main approaches can be leveraged (detailed in Appendix~B). The first approach modifies the training procedure or the underlying model architecture~\cite{grosse2017statistical,rade2021reducing,madry2018towards,sehwag2021robust}, represented by adversarial training~\cite{rade2021reducing,madry2018towards,sehwag2021robust}.
The second approach performs input transformation such as such as JPEG compression~\cite{dziugaite2016study,liu2019feature}, bit reduction~\cite{guo2017countering,xu2017feature}, pixel deflection~\cite{prakash2018deflecting}, or employing random transformations~\cite{xie2017mitigating} to eliminate the attack effect. The third approach that is mostly studied is to detect AEs~\cite{ma2018characterizing,papernot2018deep,svoboda2019peernets,feinman2017detecting,lee2018simple,cohen2020detecting,ho2022disco}.

\noindent\textbf{Limitations.}
These countermeasures all suffer one or more shortcomings. 
First, the model/training modification and input transformation often suffer an accuracy trade-off to benign examples~\cite{yang2022you,zhu2023ai,shan2020gotta,meng2017magnet}. Second, it is noted that the majority of these defenses are designed to counter AE in the imaging modality explicitly or inexplicably. They are facing challenges to be adapting to other data modalities, for example, those reconstruction based and input transformation based~\cite{yang2022you,xu2017feature,ho2022disco,xiang2022patchcleanser, wang2023addition}. Third, the majority are devised for classification tasks~\cite{yang2022you,ma2019nic,zhu2023towards}, they might be inapplicable to other tasks e.g., regression. Last, some of them require modifying the training process or model architecture~\cite{ma2019nic,zhu2023ai,shan2020gotta,zhu2023towards}, which usually comes with a substantially increased computational overhead and a trade-off of the accuracy to benign examples. As for detection methods, they can suffer a high FRR, for example, ContraNet sets a 5\% FRR for AE detection accuracy evaluations and it witnesses a detection performance drop when this tolerable FRR is set to be a bit lower~\cite{yang2022you}. In other words, it is extremely challenging to have a high detection accuracy e.g., 90\% under the constraint of a small FRR e.g., 1\%, as also evidenced in \autoref{fig:comparison}.

Notably, numerous studies depend on observed AEs, particularly those utilizing adversarial training and detection-based methods, making them intrinsically susceptible to unseen AE attacks and especially adaptive attacks~\cite{aldahdooh2022adversarial,285455}.
Thus, it becomes crucial to develop a defense mechanism that does not rely on prior knowledge of specific attacks.

\noindent\textbf{Spatial Imprint VS Temporal Imprint.}
All existing AE detection methods rely on static or spatial only adversarial imprints of the target underlying model or focus on adversarial training on the underlying model. They all overlook the information easily available from temporal trajectory imprints. we are the first to explore a new dimension: the temporal information or historical imprints of the target underlying model in the particular spectrum domain that are distinctly traceable.


\vspace{-0.3cm}
\section{Conclusion}\label{sec:conclusion}
This study has unveiled the previously elusive temporal trajectory imprinted by AE attacks, introducing a new and fresh perspective in understanding and combating adversarial behavior. Through the innovative application of synthetic loss as a universal metric, the proposed \name, adeptly identifies and isolates AEs in a diverse range of scenarios. This method stands out for its ability to encapsulate the adversarial temporal trajectory in a comprehensive and effective manner. \name's prowess has been validated across a battery of experiments, successfully countering up to 12 different types of AE attacks and strong adaptive attacks without any attacking foreknowledge. Its robustness is further underscored by its consistent performance across various model architectures, including but not limited to ResNet, ResNeXt, VGG, AudioNet, BERT, and DeepSpeech. Moreover, \name has demonstrated its versatility by performing equally well across different data modalities (such as images, audio, and text), task categories (including both classification and regression), and datasets.






%
\bibliographystyle{plain}
\bibliography{References}

\clearpage
\newpage

\appendices
\section{Adversarial Example Attack}\label{app:AE}
\noindent$\bullet$ \textbf{Image Modality.} The 7 AE attacks evaluated in \autoref{sec:evaluation} are briefly introduced, which include 6 white-box attacks and 1 black-box attack. Specifically, the white-box adversarial examples include Fast Gradient Sign Method (FGSM)~\cite{goodfellow2014explaining}, Projected Gradient Descent (PGD)~\cite{madry2017towards}, Basic Iterative
Method (BIM)~\cite{kurakin2018adversarial}, Carlini and
Wagner Attack (CW)~\cite{carlini2017towards}, DeepFool~\cite{moosavi2016deepfool}, Jacobian Saliency Map Attack (JSMA)~\cite{papernot2016limitations}. The black-box adversarial example attack is Boundary attack~\cite{brendel2017decision}. 

FGSM computes delicate perturbations to be added to the input example by leveraging the gradients of a target model to maximize loss, causing misclassification without compromising the original example semantic notably. The perturbation is injected within only one-step.
BIM also known as iterative FGSM, is an iterative variant of the FGSM, which operates by applying FGSM iteratively to generate adversarial examples. 
PGD works similarly to BIM, extending FGSM by employing multiple iterations or steps of gradient descent while projecting the perturbation onto an \(\epsilon\)-ball (a range or boundary) around the original example to ensure the perturbation remains close to the original example and within a bounded region. CW is an optimization-based adversarial attack method that formulates the AE generation as an optimization problem, aiming to find perturbations that minimize the distortion while ensuring misclassification, often considered one of the most effective and versatile AE attacks. DeepFool is an iterative adversarial attack method that computes minimal perturbations by iteratively linearizing the target model's decision boundary to find the shortest distance from an input image to misclassify it, making it highly effective in generating imperceptible adversarial perturbations. JMSA leverages the Jacobian matrix of a neural network to identify and craft adversarial perturbations by exploiting the sensitivity of the model's decision boundaries to small changes in input features, aiming to deceive the model into misclassification while considering the input's saliency information. Boundary Attack as a black-box attack starts from a randomly chosen point near the decision boundary of a victim model and iteratively moves towards the target classification boundary by performing small perturbations.

\noindent$\bullet$ \textbf{Text Modality.} To evaluate \name performance on textual modality in subsection 5.2, we consider four mainstream textual AE attacks: Projected Word-Wise Substitution (PWWS)~\cite{ren2019generating}, TextBugger~\cite{li2018textbugger}, HotFlip~\cite{ebrahimi2018hotflip}, UAT~\cite{wallace2019universal}. 
PWWS, HotFlip, and UAT are white-box attacks and TextBugger is a black-box attack.

PWWS~\cite{ren2019generating} aims to generate perturbations in text by substituting words with their semantically similar counterparts while ensuring these changes are imperceptible to human readers. Unlike PWWS, TextBugger~\cite{li2018textbugger} also makes word substitution and employs various transformations, including synonym substitution, addition, and modification, to craft adversarial examples, thereby showcasing a broader range of attack strategies in the textual modality. HotFlip~\cite{ebrahimi2018hotflip} is an adversarial attack method for text classification that operates in a white-box setting, focusing on substituting individual words in the input text to alter the classification decision of a model. It utilizes gradient information to find the most influential words to flip, aiming to generate minimal perturbations that cause misclassification. In contrast to PWWS and TextBugger, HotFlip primarily focuses on flipping words rather than employing a broader range of transformations, emphasizing a more targeted approach to crafting adversarial examples in the textual modality based on word-level perturbations and gradients. Compared to PWWS, TextBugger, and HotFlip, the strength of UAT (Universal Adversarial Triggers)~\cite{wallace2019universal} lies in its ability to create universal perturbations that have consistent adversarial effects across different inputs and models.


\begin{figure}[ht]
    \centering
    \includegraphics[trim=0 0 0 0,clip,width=0.35\textwidth]{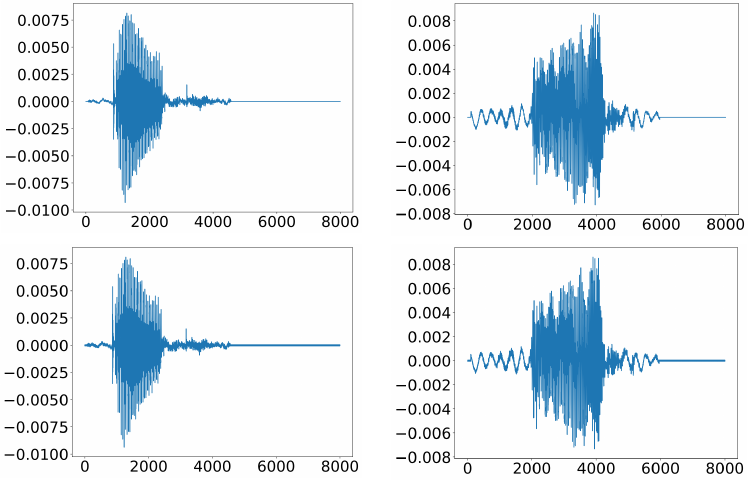}
        \caption{(Top) Benign audio waveform and (Bottom) its (Below) corresponding adversarial audio waveform. AE method of (Left) FGSM, and (Right) PGD.}
    \label{fig:audio_waveform}
\end{figure}

\begin{table}[ht]
\centering 
\caption{\name detection performance when synthetic loss computed using every two constructive IMs (CIFAR10+VGG16).}
\resizebox{0.70\linewidth}{!}
{
\begin{tabular}{cccc}
\toprule
\multirow{2}{*}{\begin{tabular}[c]{@{}c@{}}AE Attack Method\end{tabular}} &
\multirow{2}{*}{\begin{tabular}[c]{@{}c@{}}Preset FRR(\%)\end{tabular}} &
\multicolumn{2}{c}{AE Acc(\%)} \\ \cmidrule(l){3-4} & & 
\multicolumn{1}{c}{Conse.}   & Prev. \\ \midrule
\multirow{3}{*}{\begin{tabular}[c]{@{}c@{}}FGSM\\ ($\epsilon=8/255$)\end{tabular}} 
  & 1 & \multicolumn{1}{c}{98.57} & 99.01   \\  
  & 3 & \multicolumn{1}{c}{98.72} & 99.33   \\ 
  & 5 & \multicolumn{1}{c}{98.95} & 99.41   \\ \midrule
\multirow{3}{*}{\begin{tabular}[c]{@{}c@{}}PGD\\ ($\epsilon=8/255$)\end{tabular}}  
  & 1 & \multicolumn{1}{c}{98.15} & 98.79   \\ 
  & 3 & \multicolumn{1}{c}{98.48} & 99.11   \\ 
  & 5 & \multicolumn{1}{c}{98.54} & 99.24   \\ \midrule
\multirow{3}{*}{CW} 
  & 1 & \multicolumn{1}{c}{99.37} & 99.64   \\ 
  & 3 & \multicolumn{1}{c}{99.45} & 99.81   \\  
  & 5 & \multicolumn{1}{c}{99.52} & 99.86   \\ \midrule
\end{tabular}
}
\label{tab:losssys}
\end{table}

\section{AE Countermeasure}\label{app:AEcountermeasure}

To counter notorious AE attacks, three main approaches can be leveraged, which are: modifying the model and changing the training process; input transformation to eliminate the attack effect; and AE detection.

\vspace{0.2cm}
\noindent\textbf{Model/Training Modification.} Adversarial training~\cite{rade2021reducing,madry2018towards,sehwag2021robust} dominates this categorization, where AEs are included in the training data to learn features against them---regularization that changes the default training process is often applied.
This approach is bound together with the target DL model. Hence, re-training is required if the classifier changes and the cost of adversarial training increases
for a more complicated DL model. Some works require modifying the model architecture, e.g., accommodate to add an outlier class~\cite{grosse2017statistical} and then retraining the underlying model---the retraining needs knowledge of \textit{seen adversarial examples}. Besides retraining, the generation of AEs in the training exacerbates computational costs. In addition, observation of crafted AEs is a prerequisite for adversarial training, which is thus susceptible to unseen AE attacks. Moreover,  model architecture modification or training regularization could often hurt the model performance (e.g., degrading accuracy) on benign examples~\cite{grosse2017statistical,rade2021reducing}.

\noindent\textbf{Input Transformation.} 
The manipulation of input can be achieved using diverse methods such as JPEG compression~\cite{dziugaite2016study,liu2019feature}, bit reduction~\cite{guo2017countering,xu2017feature}, pixel deflection~\cite{prakash2018deflecting}, or employing random transformations~\cite{xie2017mitigating}. Beyond defenses operating within pixel space, malevolent images can also be reconstructed to better adhere to natural image statistics through the use of e.g., generative models~\cite{samangouei2018defense,zhou2023eliminating}. However, these transformations unavoidably impact benign examples, leading to a compromise between the accuracy of benign examples and the efficacy of mitigating adversarial effects. It's crucial to note that these transformations are primarily tailored for specific imaging modalities, potentially making them unsuitable or challenging to implement in other modalities. Furthermore, these methods cannot detect the AE attack, identify the attack source and the attacker, and provide no deterrence to attackers.

\noindent\textbf{Adversarial Detection.}
The exploration of input example's neighborhood information, such as local intrinsic dimensionality~\cite{ma2018characterizing}, has served as a fundamental basis in numerous studies aiming to detect AEs~\cite{papernot2018deep,svoboda2019peernets,feinman2017detecting,lee2018simple,cohen2020detecting,ho2022disco}. However, the training of the detector (referred to as the meta-classifier) in these detection studies~\cite{ma2018characterizing,abusnaina2021adversarial,papernot2018deep,svoboda2019peernets,feinman2017detecting} necessitates the observation of AEs, rendering them highly vulnerable to unseen AE attacks. Additionally, this approach can lead to significant computational overhead due to the requirement of substantial neighborhood information~\cite{abusnaina2021adversarial}.

LSD~\cite{zhang2023lsd} and NIC~\cite{ma2019nic} assess the inconsistency of a DL model's intermediate layers predicted labels to detect AEs. These defenses add internal classifiers to intermediate layers of the model e.g., CNN model, to obtain predictions from internal classifiers. ContraNet~\cite{yang2022you} utilizes a class-conditional GAN to reconstruct an incoming image based on its predicted label, leveraging semantic similarity between the reconstructed and original images to identify AEs that demonstrate lower similarity. Similar to prior work such as~\cite{xu2017feature}, it incurs a sacrifice in accuracy for benign images to some extent. Meanwhile, uGuard~\cite{bethany2023towards} concentrates on detecting unsafe images, such as those containing explicit content, within social networks. Upon detection of such images, it employs interpretability to obfuscate the harmful content. It's important to note that while~\cite{han2023interpreting} advocates for interpretability as an AE countermeasure, it remains potentially vulnerable to AE attack variant~\cite{zhang2020interpretable}.


Input preprocessing can be used to detect AEs. Feature Squeezing~\cite{xu2017feature} constructs an AE detector by checking the inconsistency of inference results when the input has undergone differing squeezing operations.
ADDITION~\cite{wang2023addition} introduces random noise into the input image and subsequently reconstructs it for denoising purposes. The denoised image is then compared with the original input image; if a prediction inconsistency beyond a certain threshold is detected, the input is deemed adversarial. However, this method proves ineffective when the adversarial perturbation is not exceedingly small while still remaining imperceptible.

Certifiably AE detection methods~\cite{shumailov2020towards,xiang2023objectseeker,xiang2022patchcleanser} stand out due to their capability of providing provable detection under assumed bounds, such as perturbation magnitude, in contrast to other approaches. However, this method often incurs heavy computational requirements and can be easily compromised, as attackers are not obliged to adhere to the underlying assumptions, such as constrained small perturbations or specific norms like $\ell_{\infty}$.

\end{document}